\documentclass[structabstract]{aa}  
\usepackage{natbib}
\usepackage{lscape}
\usepackage{rotating}
\usepackage[T1]{fontenc}
\usepackage[utf8]{inputenc} 
\usepackage[toc,page]{appendix}
\usepackage{graphicx,array}
\usepackage{enumerate}
\usepackage{amsmath}
\usepackage{subcaption}
\usepackage{tikz}
\usepackage{txfonts}
\usepackage{float}
\usepackage{textcomp}
\usepackage{gensymb}
\usepackage[colorlinks=true,linkcolor=blue,citecolor=blue]{hyperref}
\usepackage{color}
\usepackage{mathtools}
\usepackage{mathrsfs}
\usepackage[para,online,flushleft]{threeparttable}
\newcommand\devi[1]{{\textcolor{red}{#1}}}

\newcommand\MZ[1]{{\textcolor{black}{#1}}}

\begin{document}

  \title{Exploring the nature of ultra-luminous X-ray sources across stellar population ages using detailed binary evolution calculations\\}

   \author{Devina\,Misra
           \inst{1,2}\fnmsep\thanks{e-mail: devina.misra@ntnu.no},
           Konstantinos\,Kovlakas\inst{2, 3, 4},
           Tassos\,Fragos \inst{2,{5}},
           Jeff\,J.\,Andrews \inst{6},
           Simone\,S.\,Bavera \inst{2,5},
           Emmanouil\,Zapartas \inst{7,\MZ{8}},
           Zepei\,Xing \inst{2, {5}},
           Aaron\,Dotter \inst{{9}},
           Kyle\,Akira\,Rocha \inst{{9}},
           Philipp\,M.\,Srivastava \inst{9,10},
           Meng\,Sun \inst{9}
           }

   \authorrunning{Misra et al.}
    \titlerunning{Nature of ULXs across multiple stellar population ages}
   \institute{
    Institutt for Fysikk, Norwegian University of Science and Technology, Trondheim, Norway
   \and
   Département d'Astronomie, Université de Genève, Chemin Pegasi 51, CH-1290 Versoix, Switzerland
   \and
   Institute of Space Sciences (ICE, CSIC), Campus UAB, Carrer de Magrans, 08193 Barcelona, Spain
   \and 
   Institut d’Estudis Espacials de Catalunya (IEEC), Carrer Gran Capit\`a, 08034 Barcelona, Spain
   \and
    Gravitational Wave Science Center (GWSC), Université de Genève, 24 quai E. Ansermet, CH-1211 Geneva, Switzerland
   \and Department of Physics, University of Florida, 2001 Museum Rd, Gainesville, FL 32611, USA
   \and
   IAASARS, National Observatory of Athens, Vas. Pavlou and I. Metaxa, Penteli, 15236, Greece
   \and
   \MZ{Institute of Astrophysics, FORTH, N. Plastira 100,  Heraklion, 70013, Greece}
   \and Center for Interdisciplinary Exploration and Research in Astrophysics (CIERA), 1800 Sherman, Evanston, IL 60201, USA 
   \and Electrical and Computer Engineering, Northwestern University, 2145 Sheridan Road, Evanston, IL 60208, USA
     }

   \date{Received September 5, 2023; Accepted November 12, 2023}

% \abstract{}{}{}{}{} 
% 5 {} token are mandatory
 
  \abstract
    {Ultra-luminous X-ray sources (ULXs) are sources observed to have extreme X-ray luminosities exceeding the Eddington limit of a stellar-mass black hole (BH). A fraction of ULXs show X-ray pulsations, which are evidence for accreting neutron stars (NSs). Theoretical studies have suggested that NSs, rather than BHs, dominate the compact objects of intrinsic ULX populations, even though the majority of the observed sample is non-pulsating, implying that X-ray pulses from many NS ULXs are unobservable.}
   {We simulate populations of X-ray binaries covering a range of starburst ages spanning from  5 to 1000\,Myr with the aim of comparing the properties of observed ULXs at the different ages. Additionally, we compare two models describing different assumptions for the physical processes governing binary evolution.}
   {We used the new population synthesis code \texttt{POSYDON} to generate multiple populations of ULXs spanning multiple burst ages. We employed a model for geometrically beamed emission from a super-Eddington accretion disk in order to estimate the luminosities of ULXs. Following theoretical predictions for the alignment of the spin axis of an NS with the accretion disk due to mass transfer, we estimated the required mass to be accreted by the NSs in the ULX populations so that the alignment suppresses observable X-ray pulses.} 
   {While we find that the properties of ULX populations are sensitive to model assumptions, there are certain trends that the populations follow. Generally, young and old stellar populations are dominated by BH and NS accretors, respectively. The donor stars go from being massive H-rich main-sequence stars in young populations ($<100\,\rm Myr$) to low-mass post-main sequence H-rich stars in older populations ($>100\,\rm Myr$), with stripped He-rich giant donors dominating the populations at around 100\,Myr. In addition, we find that NS ULXs exhibit stronger geometrical beaming than BH ULXs, leading to an underrepresentation of NS accretors in observed populations. Coupled with our finding that X-ray pulses are suppressed in at least 60\% of the NS ULXs, we suggest that the observed fraction of ULXs with detectable X-ray pulses is very small, in agreement with observations.}
   {We show that geometrical beaming and the mass-accretion phase are critical aspects of understanding ULX observations. Our results suggest that even though most ULXs have accreting NSs, those with observable X-ray pulses would be very few.
  }


\maketitle

\section{Introduction}
Ultra-luminous X-ray sources (ULXs) are extragalactic X-ray sources with luminosities exceeding $10^{39}\, \rm erg\, s^{-1}$and are off center from their host galaxies \citep{2017ARA&A..55..303K, 2021AstBu..76....6F}. Since their first detection by the Einstein Telescope \citep[e.g.,][]{1989ARA&A..27...87F}, several hundred more ULX candidates have been observed \citep{2004ApJS..154..519S,2011MNRAS.416.1844W,2011ApJS..192...10L,2019MNRAS.483.5554E,2020MNRAS.498.4790K,2022MNRAS.512.3284S,2022A&A...659A.188B,2023MNRAS.522.1377S}. It is the general consensus that these sources are bright X-ray binaries (XRBs) with a non-degenerate star transferring mass onto a neutron star (NS) or a black hole \citep[BH;][]{2001ApJ...552L.109K, 2002ApJ...577..710Z}. The Eddington limit of a stellar-mass BH (with mass $\sim 10\,\rm M_{\odot}$) is around $10^{39}\,\rm erg\, s^{-1}$. Hence, ULXs are thought to be super-Eddington (assuming isotropic emission). The nature of the compact object (CO) involved in ULXs has been the subject of much speculation because of the apparent super-Eddington nature of ULXs. In order for the observed ULX luminosities to be explained by Eddington-limited accretion, one proposed explanation is the presence of accreting intermediate-mass BHs, that is, BHs with masses around $100$ to $1000\,\rm M_{\odot}$  \citep{1999ApJ...519...89C, 2006ASPC..352..121M, 2007Natur.445..183M}. However, it is now believed that the majority of ULXs might not belong to this class, as evidence for intermediate-mass BHs is scarce \citep[e.g.,][]{2012ApJ...755L...1M,2017MNRAS.468.2114P,2017A&A...608A..47K,2018ApJ...862...16T,2019MNRAS.482.4713Z,2019ApJ...875....1M}. Recently, it has been suggested that a ULX source (CXO\,J133815.6+043255) is an intermediate-mass BH based on study of its radio spectral energy distribution \citep{2023ApJ...956....3S}.

{X-ray binaries are categorized into three types: low-mass X-ray binaries (LMXBs; donors with masses $\lesssim 2.0\,\rm M_{\odot}$), intermediate-mass X-ray binaries (IMXBs; donors with masses in the range $2.0$ to $8.0\,\rm M_{\odot}$), and high-mass X-ray binaries (HMXBs; donors with masses $\gtrsim 8.0\,\rm M_{\odot}$). Generally, ULXs have been associated with active star-forming regions that are dominated by young and bright HMXBs \citep{2002MNRAS.337..677R,2003ApJ...596L.171G, 2004MNRAS.348L..28K, 2004A&A...426..787W, 2007ApJ...661..135Z,2016MNRAS.460.3570A,2018ApJ...863...43W, 2020MNRAS.498.4790K}. However, studies have found that intense mass-transfer phases in LMXBs and IMXBs (with accreting NSs) could produce ULX luminosities \citep{2001ApJ...552L.109K,2015ApJ...802..131S, 2018MNRAS.474.4564K, 2019ApJ...875...53W, 2019A&A...628A..19Q, 2020A&A...642A.174M}. Binaries with an expanding donor (after exhausting core hydrogen) with a radiative envelope can drive super-Eddington mass-transfer rates if the donor is slightly more massive than the accreting CO (within a certain limit so as to not initiate a dynamical instability). Also, some ULX observations have been linked to old stellar populations \citep[that are dominated by LMXBs;][]{2001ApJ...557L..35A,2002ApJS..143...25C,2004ApJ...611..846K,2004ApJS..154..519S,2006ApJ...650..879F,2008ApJ...675.1067F}.} 

The discovery of coherent X-ray pulsations (with a pulse period of $1.37$\,s) in M82\,X-2 \citep{2014Natur.514..202B} confirmed that at least a fraction of ULXs have NS accretors, as X-ray pulses imply the presence of accreting NSs. Matter accreting onto the NS surface forms a hot spot that creates X-ray pulses with the spinning of the NS \citep{1995exru.book.....S}. Since BHs do not have a well-defined surface, they cannot emit similar X-ray pulsations. Many more pulsating ULXs have been observed since M82\,X-2 \citep{2016ApJ...831L..14F, 2017MNRAS.466L..48I, 2017Sci...355..817I,2017A&A...605A..39T, 2018MNRAS.476L..45C,2018NatAs...2..312B, 2018A&A...613A..19D, 2019MNRAS.tmpL.104S,2019ApJ...879...61Z,2020ApJ...895...60R,2021MNRAS.503.5485Q}. The X-ray luminosities of pulsating ULXs are typically in the range of $10^{39}$ to $10^{41}\, \rm erg\, s^{-1}$, exceeding the NS Eddington limit, which is around $10^{38}\, \rm erg\, s^{-1}$, by a few orders of magnitude. 

There are many explanations for the apparent breach of the Eddington limit. One suggested model explaining ULXs describes the formation of an accretion disk receiving matter at super-Eddington rates \citep{1973A&A....24..337S, 1999AstL...25..508L, 2007MNRAS.377.1187P}. Within the described accretion disk, at a given disk radius, strong outflows emerge perpendicular to the disk, taking away excess matter and angular momentum and thereby keeping the disk locally Eddington-limited while the total luminosity emitted from the disk exceeds the CO Eddington limit. The structure of the accretion disk changes from a thin disk to a geometrically thick disk. The outflowing wind carries angular momentum, which creates a hollow cone through which the outgoing X-ray emission escapes. This results in an additional increase in the perceived luminosity for an observer looking face-on at this source \citep{2001ApJ...552L.109K,2009MNRAS.393L..41K}. The degree of collimation of outgoing emission depends on the nature of the accreting object. Since the Eddington limit of NSs is an order of magnitude lower than that of BHs, their X-ray emission is expected to be more beamed to reach ULX-like luminosities. Additionally, various theoretical studies suggest that a high degree of collimation of emission from geometrically thick disks can explain ULX luminosities well enough \citep[e.g.,][]{2016MNRAS.458L..10K, 2017MNRAS.470L..69M, 2019MNRAS.485.3588K, 2020MNRAS.494.3611K,2023MNRAS.526.2506L}. {Alternatively, observations of certain sources suggest that the high luminosity from ULXs could be intrinsic to the accreting CO involved. For instance, studying the pulsed X-ray emission of the ULX source RX\,J0209.6-7427 (with an accreting pulsar) during its 2019 giant outburst, \citet{2022ApJ...938..149H} estimated that the dominant source of the pulsed X-ray emission is from the fan beam of the accretion columns and that the high luminosity of the source ($\sim 1.11\times 10^{39}\rm\,erg\,s^{-1}$) is intrinsic instead of beamed.}

For accreting NSs, the presence of magnetic fields could add further complications. For instance, \citet{2019A&A...626A..18C} proposed an advection-dominated accretion disk when the matter is transferred at super-Eddington rates. The presence of strong magnetic fields would disrupt accretion disks within the magnetosphere, channeling all material onto the NS and modifying the critical mass-accretion rate. Magnetar-like fields ($\gtrsim 10^{14}\,\rm G$) have also been suggested to explain NS ULXs, as they reduce the electron scattering cross-section, leading to an increase in the Eddington limit \citep{1976MNRAS.175..395B, 2015MNRAS.454.2539M,2017MNRAS.467.1202M, 2017MNRAS.470.2799C, 2019A&A...626A..18C, 2019MNRAS.484..687M}.{Various observational studies involving the first galactic pulsating ULX, Swift J0243.6+6124, indicate the presence of multi-pole magnetic field components \citep{2020MNRAS.491.1857D,2022ApJ...933L...3K}. The magnetic field strength estimated for the source is around $1.6\times 10^{13}\rm\,G$ \citep[based on study of the cyclotron resonance scattering features, or CRSF;][]{2022ApJ...933L...3K}, which is an order of magnitude higher than that of typical accreting pulsars with a magnetic field assumed to have only dipole components. \citet{2022ApJ...938..149H} studied the pulsed emission from the ULX RX\,J0209.6-7427 and estimated the magnetic field strength to be about 4.8 to $ 8.6\times 10^{12}\rm\,G$ or 1.7 to $2.2\times 10^{13}\rm\,G$, both estimates correspond to dipole and multi-pole magnetic fields of the NS, respectively. A similar result was reached by \citet{,2022MNRAS.517.3354L} for this source and for SMC\,X-3 during its outburst. However, CRSF} studies of some other ULXs have suggested pulsar-like magnetic fields in the range of $10^{11}$ to $10^{13}\,\rm G$, which is below the magnetic field strength required for magnetars \citep{2018ApJ...856..128W, 2018NatAs...2..312B, 2019A&A...621A.118K}. Recently, \citet{2023MNRAS.526.2506L} suggested that magnetar-like field strengths cannot explain ULX spin-up rates and inferred that beaming due to the accretion disk must be involved in ULXs.

While, observationally, non-pulsating ULXs dominate (so far) over pulsating ULXs in number, with only about ten of the more than 500 ULX candidates known to pulsate, the absence of observed X-ray pulsations does not negate the presence of an accreting NS. Hence, many more ULXs may contain NS accretors instead of BHs, and their X-ray pulses may simply not have been observed. This argument is supported by the fact that a majority of LMXBs (which are thought to contain accreting NSs) do not show measurable X-ray pulsations \citep{1994ApJ...435..362V, 2005ApJ...626..333D,2015ApJ...806..261M,2018ApJ...859..112P}. There is also a source (namely, M51\,ULX-8) confirmed to be a ULX by observed cyclotron resonance scattering features for which X-ray pulses have not been observed so far. The observed scattering features suggest the presence of an NS with a magnetic field of around $10^{12}\,\rm G$ \citep{2018NatAs...2..312B, 2019MNRAS.486....2M}.

\citet{2017Sci...355..817I} discovered a pulsating ULX in NGC\,5907, which is also one of the most luminous ULXs known at an X-ray luminosity of $10^{41}\, \rm erg\, s^{-1}$, suggesting that other extremely bright ULXs might have accreting NSs. \citet{2017ApJ...836..113P} studied the spectral properties of bright ULXs and concluded that most ULXs fit well with the X-ray spectra of accreting magnetic NSs in the Galaxy. In fact, many theoretical studies have suggested that NSs dominate over BHs as accretors in the total ULX population, and the low number of confirmed NS ULXs is an observational effect \citep[e.g.,][]{2015MNRAS.454.2539M, 2016MNRAS.458L..10K, 2017MNRAS.468L..59K, 2017MNRAS.470L..69M, 2020MNRAS.494.3611K}. \citet{2018ApJ...856..128W} found spectral signatures in a large sample of bright ULXs (all so far non-pulsating) that are usually associated with pulsar-like emissions. \citet{2015ApJ...810...20W} studied the evolutionary channels that lead to extreme mass-transfer phases in close XRBs using binary population synthesis. In their study, of the brightest ULXs that exceeded $10^{42}\, \rm erg\, s^{-1}$ (these sources are also known as hyperluminous X-ray sources, or HLXs), half contained accreting NSs instead of BHs. While the authors focused on HLXs, the $50\%$ contribution from NS XRBs to HLXs contrasts with the expectation that BH XRBs are always brighter than NS XRBs.

\citet{2016MNRAS.458L..10K, 2020MNRAS.494.3611K} suggested that NS ULXs and BH ULXs are indistinguishable at some point during their individual evolution. Accretion would eventually align the NS spin and the accretion disk axes, obscuring the X-ray pulses. They also found that NS ULXs, when observed, would be brighter than BH ULXs, as they have stronger beaming due to their lower Eddington limit. This also suggests why HLXs are expected to have a significant contribution from NS ULXs. However, short durations of the intense mass-transfer phase and increased beaming would down-weight their observability. When studying ULX populations by taking into account their observability due to geometrically beamed emission, \citet{2019ApJ...875...53W} also found that the majority of their simulated NS ULXs had beamed emission with the mass-transfer phase proceeding on a thermal timescale compared to BH ULXs (which were typically isotropically emitting and proceeded on a nuclear timescale). They found that NS ULXs dominated the total underlying population of simulated ULXs in populations with constant star formation and that the observed ULX population was populated by NS ULXs and BH ULXs equally. For starburst populations, they found that BH ULXs dominated both observed and total populations when the systems are young ($\lesssim 1000\,\rm Myr$) but that this dominance shifts to NS ULXs for older systems. As a consequence of geometrically beamed emission in NS ULXs, observations from flux-limited surveys would be dominated mostly by BH ULXs.

The relative importance of NSs versus BHs in the intrinsic ULX populations depends not only on their accretion physics but also on the properties of the host galaxy, including age and metallicity. \citet{2015ApJ...802..131S} used a combination of parametric population synthesis calculations \cite[modified \texttt{BPS} code;][]{2002MNRAS.329..897H, 2014ApJ...796...37S} with detailed binary evolution models \cite[\texttt{TWIN} version of the \texttt{Eggleton} code;][]{1971MNRAS.151..351E, 1972MNRAS.156..361E} and found that accreting NSs dominate ULX populations over BHs in galaxies like M82 and the Milky Way when comparing relatively young and old populations, respectively, with constant star-formation rates (SFRs). \citet{2017ApJ...846...17W} found that BHs dominate only at early times in starbursts, while NSs dominate at later times in starbursts as well as in constant star-formation scenarios (for solar metallicity regions). However, for galaxies with \devi{a} constant SFR in sub-solar metallicity environments, BH ULXs were still more abundant, as a lower metallicity leads to more compact, massive stellar cores that collapse to more massive COs. \citet{2019ApJ...875...53W} found that young populations (with stellar ages $\lesssim 10\,\rm Myr$) that were dominated by BH ULXs had more beamed BH ULXs than in older populations, while the majority of NS ULXs were beamed at all epochs. Consequently, the question of the dominant type of accretor involves the considerations of many assumptions that affect the underlying stellar populations. 

In this work, we study populations of ULXs formed at starbursts of different ages, investigating the effects of age on ULXs. We also compare two sets of model assumptions that differ in certain physical properties, such as the natal kick velocities. Starburst populations give us valuable insights into the dependence of ULXs on the CO accretion and on the age of the stellar population and how these dependencies affect the sub-populations of NS ULXs and BH ULXs. Alternatively to synthetic starburst populations, simulations with continuous star-formation scenarios give us information about the general properties of ULX populations in large samples of galaxies while obscuring the effects of age on the populations. To carry out this study, we used the newly developed binary population synthesis code \texttt{POSYDON} \citep{2023ApJS..264...45F}, an open-source framework that allows for population studies where the entire life of a binary is modeled using detailed and self-consistent stellar structure and binary evolution calculations. Our ULX population models are based on the earlier study by \citet{2023A&A...672A..99M}, who modeled the X-ray luminosity function of extragalactic populations of HMXBs. Even though more ULXs per SFR have been observed in lower than solar metallicity environments \citep[for instance,][]{2013ApJ...769...92P,2016ApJ...818..140B, 2020MNRAS.498.4790K}, the majority of observed ULXs are in solar metallicity environments \citep{2020MNRAS.498.4790K}. Hence, we limited this study to solar metallicity \citep[$Z_{\odot}=0.0142$;][]{2009ARA&A..47..481A}. 

In Section\,\ref{sec:numericaltools} we discuss the population synthesis code we employed and present the model assumptions used to carry out the population synthesis study and the calculations of X-ray luminosities of accreting COs. In Section\,\ref{sec:results}, we present the various burst populations corresponding to a range of ages spanning from 5 to 1000\,Myr. We discuss the demographics of the ULX populations at various ages, including the nature of the accretors and the donors, and explore the effect of accretion on observable X-ray pulses. In Section\,\ref{sec:disc_and_conclusions}, we compare our results to those in the literature and present our concluding remarks. 

\section{Methods}
\label{sec:numericaltools}

We used \texttt{POSYDON} \citep[][ code version v1]{2023ApJS..264...45F}, a newly developed binary population synthesis code that incorporates detailed stellar structure and binary evolution tracks computed using the stellar evolution code Modules for Experiments in Stellar Astrophysics {\citep[code version 11701;][]{2011ApJS..192....3P,2013ApJS..208....4P,2015ApJS..220...15P,2018ApJS..234...34P,2019ApJS..243...10P,2023ApJS..265...15J}}. Accurate information about the structure of the stars via the detailed evolutionary sequences is expected to lead to a physically accurate estimation of the mass-transfer rate stability (particularly for thermal timescale mass-transfer), stellar rotation, and transport of angular momentum within and between the binary stars. Since all the previous studies of synthetic populations of NS and BH ULXs have been carried out, at least in part, using parametric population codes, it is imperative to study these populations by including information from detailed stellar structure calculations. Parametric codes, or rapid population codes, approximate the stellar structure using fitting formulas \citep[e.g., BSE; ][]{2000MNRAS.315..543H, 2002MNRAS.329..897H} or look-up tables \citep[e.g., ComBine;][]{2018MNRAS.481.1908K} constructed from single stellar evolutionary tracks that can introduce systematic biases when binary interaction is involved.

\subsection{Parameters of models used}\label{sec:model_params}

In this study, we focus on the sub-population dominating ULX populations that are young, bright HMXBs (with high mass-transfer rates), and found in active star-forming galaxies. \citep{2002MNRAS.337..677R,2003ApJ...596L.171G, 2004MNRAS.348L..28K, 2004A&A...426..787W, 2007ApJ...661..135Z,2016MNRAS.460.3570A,2018ApJ...863...43W, 2020MNRAS.498.4790K}. \citet{2023A&A...672A..99M} studied the effects of different combinations of physics parameters on the synthetic X-ray luminosity function (XLF) of HMXBs and found models that best matched the observed XLFs of HMXBs from star-forming galaxies \citep[taken from][]{2019ApJS..243....3L}. We took two models from that study, which we indicate as A and B \citep[64 and 44, respectively, in][]{2023A&A...672A..99M}, with the model parameters described by Table\,\ref{table:c4_best_fit}. These models correspond to the ones that best fit the observations from two different approaches: first, matching the slope of the XLFs (here model\,A) and, secondly, matching the normalization of the XLFs (here model\,B). 

The primary difference between the two models is in the normalization of the supernova (SN) kicks during the formation of a BH. In model\,A, BH kicks are normalized using the mass of the newly formed BH, while in model\,B, the BH kicks are not normalized, and they receive the same kick velocities as NSs (drawn from a Maxwellian distribution with a velocity dispersion of $265\,\rm km\,s^{-1}$). The strengths of the kicks have an observable effect on XRB populations. Stronger kicks, similar to kicks received by NS binaries, result in the disruption of wide binaries (with orbital periods $\gtrsim 100\,\rm days$) during the SN event \citep{2023A&A...672A..99M}. Another difference between the two models is in the efficiency of the common-envelope (CE) phase, or $\alpha_{\rm CE}$, which is 1.0 for model\,A (corresponding to full orbital energy being available for envelope ejection) and 0.3 (corresponding to 30\% of the orbital energy being available for envelope ejection) for model\,B. The $\alpha_{\rm CE}$ parameter affects XRBs with He-rich donors, with lower efficiency leading to a higher merger rate of binaries in the CE phase. The formation of most XRBs, however, especially those that have H-rich donors and BH accretors, do not go through a CE, and hence the $\alpha_{\rm CE}$ parameter does not affect their evolution. As we will see later, XRBs with He-rich donors that consist of accreting NSs are affected by $\alpha_{\rm CE}$, and this effect is more apparent when NSs start to dominate the ULX populations. Finally, there is an additional difference between the models, namely, in the observability of wind-fed accreting binaries. However, this parameter affects the XRBs at luminosities lower than $10^{38}\, \rm erg\, s^{-1}$ and is hence below our limit for ULXs and has a negligible effect in our study.

\begin{table*}[]
\centering
\begin{tabular}{l|llll}
\hline\hline
Parameters     & Model A   & Model B                                                                             \\ \hline
Remnant mass prescription   &  \citet{2020MNRAS.499.2803P} & \citet{2020MNRAS.499.2803P} \\
Natal kick normalization   & BH mass normalized kicks   & No kick normalization            \\
Orbit circularization at RLO & Conserved angular momentum    & Conserved angular momentum  \\
CE efficiency ($\alpha_{\rm CE}$). & 1.0        & 0.3    
\\
CE core-envelope boundary    & At $X_{\rm H} = 0.30$ & At $X_{\rm H} = 0.30$              \\
Observable wind-fed disk          &  \citet{2021PASA...38...56H} & No criterion  
\\\hline
\end{tabular}
\caption{Physical parameters corresponding to the best-fitting models\,A and B from \citet{2023A&A...672A..99M}. }
\label{table:c4_best_fit}
\end{table*}

\subsection{Properties of the initial binary population}

In order to study the effect of the age of the host population on the properties of ULXs, we simulated burst populations of $10^7$ binaries, initially at zero-age main-sequence (ZAMS), with the range of burst ages being 5\,Myr, 10\,Myr, 40\,Myr, 100\,Myr, 300\,Myr, and 1000\,Myr. The reason we chose this range of burst ages is because theoretical studies of burst populations using rapid population codes have shown that variations in ULX demographics cease beyond $\sim 1\,\rm Gyr$, which is when the populations are composed of only NS ULXs. From this point onward, these NS ULXs only decrease in number as their donors reach the end of their evolutionary lifetimes \citep{2017ApJ...846...17W,2019ApJ...875...53W}. Each of our burst populations averages at around $2.7\times 10^{8}\rm\,M_{\odot}$, and for the same initial total population mass, synthetic populations with younger burst ages sample binaries for a Milky Way-like galaxy better (which would have an SFR of about 54\,M$_{\odot}\,\rm yr^{-1}$ for a starburst age of 5\,Myr) as compared to older burst ages (which would have an SFR of about 0.27\,M$_{\odot}\,\rm yr^{-1}$ for a starburst age of 1000\,Myr). Over recent history, the SFR of the Milky Way is estimated to be less than 2\,M$_{\odot}\,\rm yr^{-1}$ \citep{2011AJ....142..197C,2015ApJ...806...96L}. Hence, the generated synthetic populations sufficiently sample the young stellar populations in star-forming regions where ULXs are prominent. 

The initial binaries were generated by drawing the primary stellar mass from the Kroupa initial mass function \citep{2001MNRAS.322..231K} in the mass range [7.0, 120.0]\,M$_{\odot}$ and the secondary mass from a uniform mass-ratio distribution \citep{2013A&A...550A.107S}. The initial orbits were taken to be circular, with periods drawn from \citet{2013A&A...550A.107S} and with the same range as the \texttt{POSYDON} grids from 0.35 to $10^{3.5}\,\rm days$ \citep[extrapolated down to 0.35, as the distribution from \citealp{2013A&A...550A.107S} is limited to 1\,day;][]{2023ApJS..264...45F}. We normalized the populations using a correction factor of $\approx 5.89$ \citep[including a binary mass function of 0.7;][]{2012Sci...337..444S} to account for the unsampled region of the initial mass function, 0.08 to 7.0\,M$_{\odot}$ \citep{2020A&A...635A..97B}. As mentioned previously, all the simulations were carried out at solar metallicity \citep[0.0142;][]{2009ARA&A..47..481A}. 

% mass accretion luminosity = RLO + wind
\subsection{X-ray luminosity calculation}
\label{sec:methods:x-ray_lum}
The X-rays from an XRB are generated from stellar material being captured by the CO. There are three types of mass-transfer phases in XRBs: mass transfer from the inner Lagrangian point in semi-detached Roche lobe overflowing binaries, mass transfer from stellar winds leaving the donor surface that are captured by the gravitational pull of the accretor, and mass transfer from a decretion disk of the donor (which is a highly-spinning B star, i.e., a Be star) when the CO interacts with the disk in an eccentric orbit. Roche-lobe overflow (RLO) occurs when the donor star starts to fill out the volume of the gravitational equipotential surface passing through the inner Lagrangian point, and material leaves the donor surface accreting onto the CO companion. Massive stars tend to have very high rates of stellar-wind loss, a fraction of which gets captured by the accretor. To account for wind mass transfer, we used the mechanism described by \citet{1944MNRAS.104..273B}. 

The third type of mass transfer for Be XRBs is treated separately. Since the modeling of the decretion disk was not carried out in \texttt{POSYDON}, we identified Be XRBs and assigned them X-ray luminosities \citep[using the same treatment as][]{2023A&A...672A..99M}. \citet{2020ApJ...902..125A} carried out population synthesis studies for a binary population with a constant SFR of 3\,M$_{\odot}\, \rm yr^{-1}$, including the effect of geometrical beaming of the X-ray emission. They found that pulsating ULXs tend to have Be- or intermediate-mass donors. The Be XRB luminosities in our populations reach up to $\sim 10^{38}\, \rm erg\, s^{-1}$, and none appear as ULXs \citep[see ][]{2023A&A...672A..99M}. While there are a few Be XRBs that have been observed to have luminosities reaching $\sim 10^{39}\, \rm erg\, s^{-1}$ during outbursts \citep{1982Natur.297..568S, 2018ApJ...863....9W, 2019ApJ...873...19T, 2020MNRAS.494.5350V, 2020MNRAS.495.2664C}, so far no Be XRBs have been observed to be persistently emitting ULX-like luminosities.

%Spinning B stars in Be XRBs have a decretion disk around them which could start mass transfer if it reaches the Roche-lobe radius of the donor. Since the modeling of the decretion disk is not carried out in \texttt{POSYDON}, we identify Be XRBs and assign X-ray luminosities to them. We use the same criteria as \citep{2023A&A...672A..99M} to identify Be XRBs, and that is wide detached binaries (orbital periods in the range of 10 to 300 days) with fast-spinning hydrogen MS donors ($\gtrsim 70\%$ of critical surface velocity), with the exception of the lower limit of Be-star masses, for which we use $6.0\,\rm M_{\odot}$ \citep{2010AN....331..349H}, while \citet{2009ApJ...707..870B} and \citet{2014MNRAS.437.1187Z} use $3.0\, \rm M_{\odot}$. We also consider the decretion disk radius \citep[using an approximated radius of $100\,\rm R_{*}$ with $R_{*}$ being the stellar radius;][]{2017A&A...601A..74K} exceeding the donor Roche-lobe radius at the periastron as an identifying condition \citep[see also][]{2023A&A...672A..99M}. We estimate the X-ray luminosities of Be XRBs using an empirical relation based on peak X-ray luminosities of the observations of the Be XRBs \citep{2006ApJ...653.1410D}.

When the mass-transfer rate ($\dot{M}_{\rm tr}$) is below the Eddington limit, mass accretion onto the CO is assumed to be fully conservative: All the mass lost by the donor via RLO is accreted onto the CO. We used the following equation to estimate the X-ray luminosity for sub-Eddington mass-transfer rates, including RLO and wind-fed accretion,
% \begin{equation}
%     L^{\rm RLO/wind}_{\rm X} = \eta \dot{M}_{\rm acc} c^2,
% \end{equation}
\begin{equation}
    L^{\rm RLO/wind}_{\rm X} = \eta \dot{M}_{\rm acc} c^2, \ \ \ \ \text{if } \dot{m} \leq 1.0,
\end{equation}
where $\dot{M}_{\rm acc}$ is the mass-accretion rate, $\eta$ is the radiative efficiency of accretion, $c$ is the speed of light, and $\dot{m}\equiv \dot{M}_{\rm tr}/\dot{M}_{\rm Edd}$ is the Eddington ratio. The radiative efficiency of accretion is estimated by using the properties of the accretor,
\begin{equation}
    \eta = \frac{GM_{\rm acc}}{R_{\rm acc} c^2},
\end{equation}
where $M_{\rm acc}$ and $R_{\rm acc}$ are the mass and radius of the accretor \citep[for BHs, $R_{\rm acc}$ is the spin-dependent innermost stable circular orbit;][]{2003MNRAS.341..385P}, and $G$ is the gravitational constant.

As the mass-transfer rate approaches and exceeds the Eddington limit, mass accretion onto the CO is limited by the Eddington limit ($L_{\rm Edd}$), and we followed the accretion disk model by \citet{1973A&A....24..337S} in order to model the accretion disk that receives material at a super-Eddington rate. This was combined with the prescription from \citet{2001ApJ...552L.109K} and \citet{2009MNRAS.393L..41K} for the additional effect of geometrical beaming. The isotropic-equivalent observed X-ray luminosity is defined as follows,
% \begin{equation}\label{eq:c4_superedd}
%     L^{\rm RLO/wind}_{\mathrm{X, iso}} = \frac{L_{\mathrm{Edd}}}{b}(1 + \ln{\dot{m}}),
% \end{equation}
\begin{equation}\label{eq:superedd}
    L^{\rm RLO/wind}_{\mathrm{X, isotropic}} = \frac{L_{\mathrm{Edd}}}{b}(1 + \ln{\dot{m}}), \ \ \ \ \text{if } \dot{m} > 1.0,
\end{equation}
where $\dot{m}$ is the ratio of the mass-transfer rate and the Eddington rate, and $b$ is the beaming factor reflecting the geometrical collimation of the emission from the thick disk. The approximate value of $b$ is given by \citet{2009MNRAS.393L..41K},
\begin{equation}\label{eq:c4_beaming}
        b= 
\begin{dcases}
    \frac{73}{\dot{m}^2},& \text{if } \dot{m}> 8.5,\\
    1,              & \text{otherwise}.
\end{dcases}
\end{equation}
For very high mass-transfer rates, the prescription above might lead to extremely strong beaming ($b \ll 10^{-3}$). We followed \citet{2017ApJ...846...17W} in setting a lower limit for $b$ at $3.2\times 10^{-3}$, which approximately corresponds to an opening angle of $9\degree$. The beaming factor was used to down-weight all the synthetic XLFs presented in this work. While geometrical beaming reduces the number of binaries observed, it also leads to an increase in the luminosity of the brightest ULXs, as the binaries are more strongly beamed.

\subsection{Suppression of pulses caused by accretion onto a neutron star}
\label{sec:methods:supression_pulses}
There is at least one confirmed NS ULX that does not (yet) show X-ray pulsations, and that is M51 ULX-8 \citep{2018NatAs...2..312B, 2019MNRAS.486....2M}. \citet{2017MNRAS.467.1202M} suggested the lack of observed X-ray pulses is caused by an optically thick envelope that smears and obscures them. The reprocessing of the emitted hard radiation turns it into a blackbody-like emission, which depends on the optical thickness of the envelope and makes the cyclotron scattering lines disappear (which is characteristic of magnetic NSs). The observation of pulses, therefore, depends on the viewing angle with respect to the NS spin and the envelope. 

\citet{2020MNRAS.494.3611K} postulated that the observations of pulses in accreting NSs are due to the misalignment of the NS spin axis and the beaming axis (refer to Fig.\,2 in their paper for more clarity). They showed that sinusoidal pulsed light curves are created when either of two conditions is achieved: (1) the spin and accretion disk axes are strongly misaligned and the magnetic axis is aligned close to the spin, or (2) the spin and disk axes are strongly aligned and the magnetic axis is misaligned with the spin. For both of these configurations, the escaping pulses are different. If the spin and disk axes are strongly misaligned, pulses escape the beaming tunnel (as the magnetic axis, and hence the hot spot, follow the spin axis) and can be observed. However, accretion tends to rapidly align the spin axis of an NS with the orbital axis. When aligned, as pulsed light curves are created when the magnetic axis is misaligned with the NS spin, most pulses are scattered by the beaming tunnel and pulsations are negligible.

The magnetic field of an NS can affect its accretion phase if it is strong enough (about $\gtrsim 10^{12}\,\rm G$). A strong magnetic field has a large Alfv{\'e}n radius, allowing the accreted angular momentum to efficiently spin up the NS. However, a weak magnetic field allows the accretion disk to extend all the way to the NS surface. To investigate how accretion-induced alignment of the NS axes would result in observable pulses, we performed an order of magnitude calculation of the angular momentum of the incoming matter ($J_{\rm acc}$) at the magnetosphere radius ($R_{\rm M}$) and at the NS surface ($R_{\rm NS}$) and equated it to the angular momentum of a fiducial NS ($J_{\rm NS}$) in order to find the amount of mass that should be accreted for the pulses to be buried,
\begin{equation}\label{eq:am_accreted}
    J_{\rm acc} = J_{\rm NS} \Leftrightarrow I_{\rm acc} \omega_{\rm acc} = I_{\rm NS} \omega_{\rm NS},
\end{equation}
where $I_{\rm NS}$ is the accretor moment of inertia and $\omega_{\rm NS}$ is the accretor angular velocity. When the condition in Equation\,\ref{eq:am_accreted} is fulfilled, the NS is assumed to have gained enough angular momentum to align its spin and orbit axes, and there will be no observable pulses. Using the Keplerian velocity for the velocity of matter at radius $r$, we got $\omega_{\rm acc} = \sqrt{GM_{\rm NS}/r^3}$ and the moment of inertia for accreted matter $\Delta m$ as $I_{\rm acc} = \Delta m r^{2}$. Thus, we arrived at an equation describing the accreted mass,
\begin{equation}\label{eq:delta_m_eq}
    \Delta m = \frac{I_{\rm NS}\omega_{\rm NS}}{\sqrt{GM_{\rm NS}r}}.
\end{equation}
%Simplifying the above equation for $\Delta m$,
% \begin{equation}
%     \Delta m = 5.46 \times 10^{35} \times \bigg(\frac{M_{\rm NS}}{\rm M_{\odot}}\bigg)^{-1/2} \bigg(\frac{r}{1 \rm cm}\bigg)^{-1/2} \rm g.
% \end{equation}

We assumed $I_{\rm NS}=10^{45}\,\mathrm{g}\,\mathrm{cm}^2$ (approximate value for a 1.3\,M$_{\odot}\,$NS) and took $\omega_{\rm NS}=2\pi / P_{\rm spin}$, with $P_{\rm spin}=1$s \citep[similar to the spin period of the most well-studied pulsating ULX M82\,X-2, which is $1.37$\,s;][]{2014Natur.514..202B}. For the matter being accreted at the surface of the NS \citep[hence, r is taken to be the NS radius, which is R$_{\rm NS}$ = 12.5\,km;][]{2018PhRvL.120z1103M, 2019ApJ...887L..24M, 2019ApJ...887L..21R, 2020PhRvD.101l3007L, 2020ApJ...892L...3A, 2021A&A...650A.139K, 2021ApJ...921...63B, 2021ApJ...918L..29R}, we could get an estimate for the amount of matter that needs to be accreted for pulses to be unobservable,
% \begin{equation}\label{eq:delta_m_Rns}
%      \Delta m (R_{\rm NS}) = 2.45\times 10^{-4} \bigg(\frac{M_{\rm NS}}{1\rm M_{\odot}}\bigg)^{-1/2} \bigg(\frac{R_{\rm NS}}{12.5\rm km}\bigg)^{-1/2} \bigg(\frac{P_{\rm spin}}{1\,\rm s}\bigg)^{-1} \rm M_{\odot}.
% \end{equation}
\begin{equation}\label{eq:delta_m_Rns}
     \Delta m (R_{\rm NS}) = 2.07\times 10^{-4} \bigg(\frac{M_{\rm NS}}{1.4\rm M_{\odot}}\bigg)^{-1/2} \bigg(\frac{R_{\rm NS}}{12.5\rm km}\bigg)^{-1/2} \bigg(\frac{P_{\rm spin}}{1\,\rm s}\bigg)^{-1} \rm M_{\odot}.
\end{equation}
The case described above would correspond to the case when there is a weak magnetic field, and the accretion disk can be approximated to extend down to the NS surface. In an NS with a strong magnetic field, the magnetosphere radius is defined as the radius where the magnetic stresses dominate the accretion flow \citep{frank2002accretion}, and it is defined as
\begin{equation}\label{eq:rm_radius}
    R_{\rm M} = (2GM_{\rm NS})^{-1/7} \dot{M}_{\rm NS}^{-2/7} \mu^{4/7},
\end{equation}
% \citet{2016MNRAS.458L..10K} approximated the magnetosphere radius using the description of accretion onto a magnetized NS done as follows,
% \begin{equation}\label{rm_king}
%     R_{\rm M} = 2.9 \times 10^8 \times \bigg(\frac{\dot{M}_{\rm donor}}{10^{17}\rm g\, s^{-1}}\bigg)^{-2/7} \bigg(\frac{M_{\rm NS}}{1.4 \rm M_{\odot}}\bigg)^{-1/7} \bigg(\frac{\mu}{10^{30}\rm G\,cm^3}\bigg)^{4/7} \rm cm.
% \end{equation}

where $\mu$ is the NS magnetic dipole moment (which has  a typical value of $10^{30}\,\rm G\,\rm cm^3$ that  corresponds to an NS surface magnetic field strength of $B_{\rm NS} \sim \mu / \rm R_{NS}^{3} \approx 10^{12}\,\rm G$). By simplifying Equation\,\ref{eq:rm_radius} and rewriting it in convenient units,
we obtained% \begin{equation}\label{eq:rm_radius_units}
%     \nonumber R_{\rm M} = 1.916\times 10^{8}  \bigg(\frac{M_{\rm NS}}{1\rm M_{\odot}} \bigg)^{-1/7}  \bigg(\frac{\dot{M}_{\rm NS}}{10^{-8}\rm M_{\odot} yr^{-1}} \bigg)^{-2/7} \\ 
%     \bigg(\frac{\mu}{10^{30}\,\rm G\, cm^{2}}\bigg)^{4/7} \rm cm. 
% \end{equation}
\begin{equation}\label{eq:rm_radius_units}
    \nonumber R_{\rm M} = 1.83\times 10^{8}  \bigg(\frac{M_{\rm NS}}{1.4\rm M_{\odot}} \bigg)^{-1/7}  \bigg(\frac{\dot{M}_{\rm NS}}{10^{-8}\rm M_{\odot} yr^{-1}} \bigg)^{-2/7} \\ 
    \bigg(\frac{\mu}{10^{30}\,\rm G\, cm^{2}}\bigg)^{4/7} \rm cm. 
\end{equation}
Using Equations\,\ref{eq:rm_radius_units} and \ref{eq:delta_m_eq}, we could then estimate the mass required to spin up the NS to $P_{\rm{spin}}$ if all the matter at $R_{\rm M}$ is accreted due to the magnetic field of the NS,
% \begin{align}\label{eq:delta_m_Rm}
%     \nonumber \Delta m (R_{\rm M}) = 1.98\times 10^{-5} \bigg(\frac{M_{\rm NS}}{1\rm M_{\odot}}\bigg)^{-3/7} \bigg(\frac{\dot{M}_{\rm NS}}{10^{-8}\rm M_{\odot} yr^{-1}} \bigg)^{1/7} \\
%     \bigg(\frac{\mu}{10^{30}\,\rm G\, cm^{2}}\bigg)^{-2/7} \bigg(\frac{P_{\rm spin}}{1\,\rm s}\bigg)^{-1}  \rm M_{\odot}.
% \end{align}
\begin{align}\label{eq:delta_m_Rm}
    \nonumber \Delta m (R_{\rm M}) = 1.71\times 10^{-5} \bigg(\frac{M_{\rm NS}}{1.4\rm M_{\odot}}\bigg)^{-3/7} \bigg(\frac{\dot{M}_{\rm NS}}{10^{-8}\rm M_{\odot} yr^{-1}} \bigg)^{1/7} \\
    \bigg(\frac{\mu}{10^{30}\,\rm G\, cm^{2}}\bigg)^{-2/7} \bigg(\frac{P_{\rm spin}}{1\,\rm s}\bigg)^{-1}  \rm M_{\odot}.
\end{align}

After the NS has accreted enough matter so that the axis of the NS will be aligned with the disk axis, pulsations will not be observed. Hence, we used Equations\,\ref{eq:delta_m_Rns} and \ref{eq:delta_m_Rm} to identify ULXs with NSs that have not yet accreted enough matter in order to suppress the emitted X-ray pulses. The default populations presented in our work do not automatically take into account the suppression of beamed emission in the detectability of NS ULXs. We present the results including this effect separately in Section\,\ref{sec:results:sup_pulse}.

We note that the presented model assumes that the X-ray radiation comes from the hot spot created on the NS surface that follows the magnetic fields lines. If the emission is from the walls of the accretion column with a fan-beamed pattern, as postulated for the ULX source RX\,J0209.6-7427 \citep{2022ApJ...938..149H}, the assumed model might not be accurate. In this case, the inferred accretion rate is also super-Eddington. Since the deciding factor in the aligning of the spin and beaming axes is the amount accreted by the NS (defined by Equations\,\ref{eq:delta_m_eq} and \ref{eq:delta_m_Rm}), for accretion rates higher than the standard scenario, the alignment would be much quicker and even fewer NS ULXs would be seen.

\begin{figure*}[!ht]
\centering
\includegraphics[width=0.8\linewidth]{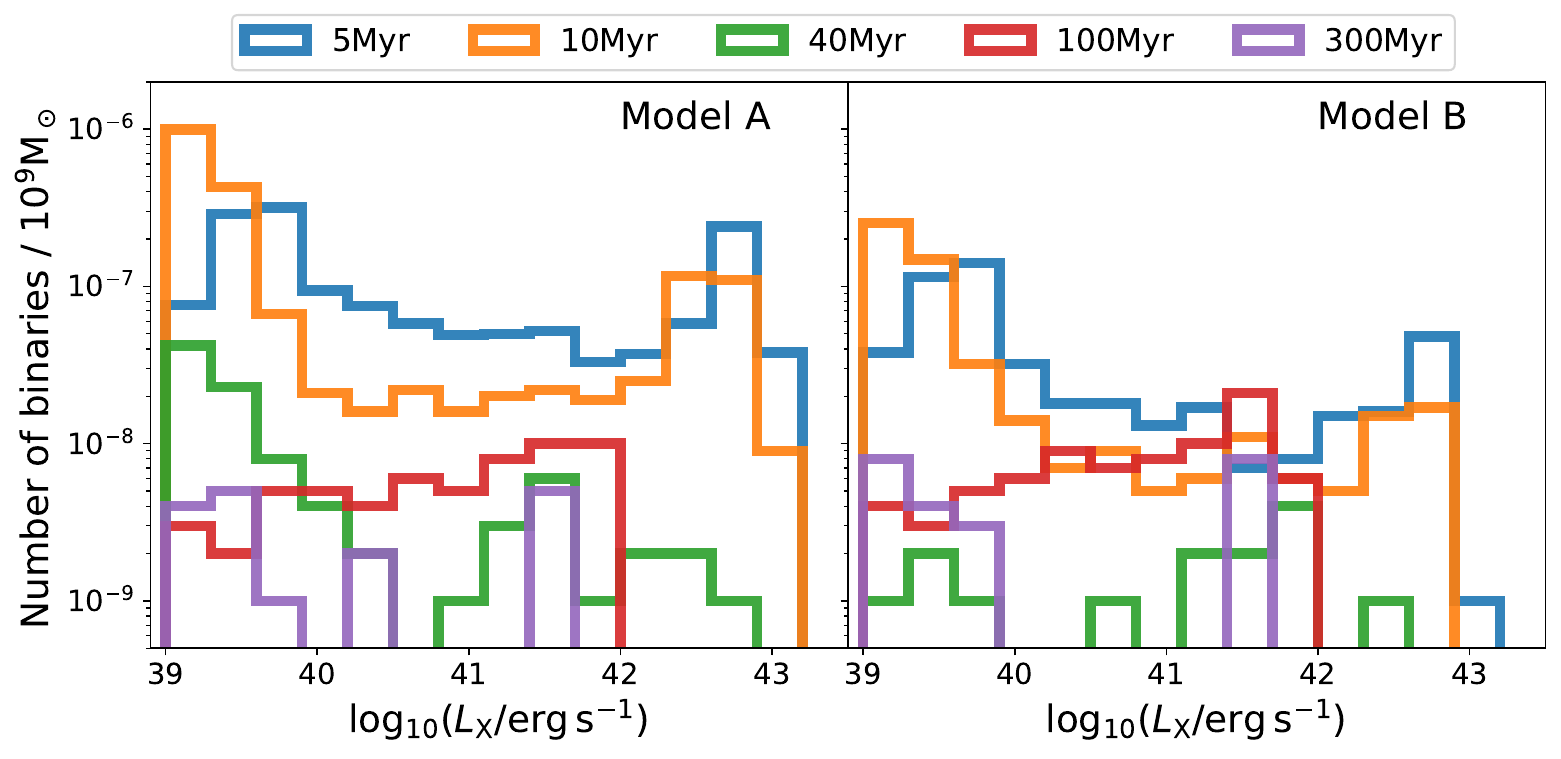}
\caption{Distribution of X-ray luminosities of ULXs for all the burst ages (see legend) for the populations following models\,A and B (left and right panel, respectively). Geometrical beaming is accounted for in the X-ray luminosity, but the observability of a ULX is not. In the observed population, the brighter luminosities would be slightly suppressed, as they would be beamed.}
\label{fig:xrb_number}
\end{figure*}

\begin{figure*}[htb]
\hspace{2.0cm}\includegraphics[width=0.7\textwidth]{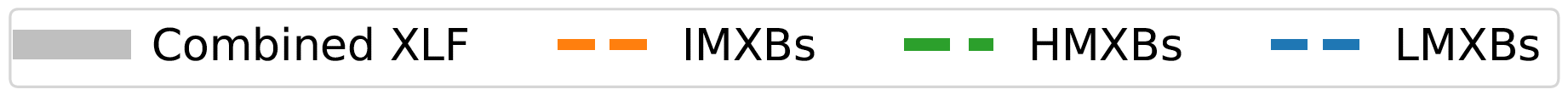}\\
\centering
\begin{tikzpicture}
\node (img1)  {\includegraphics[width=0.95\textwidth]{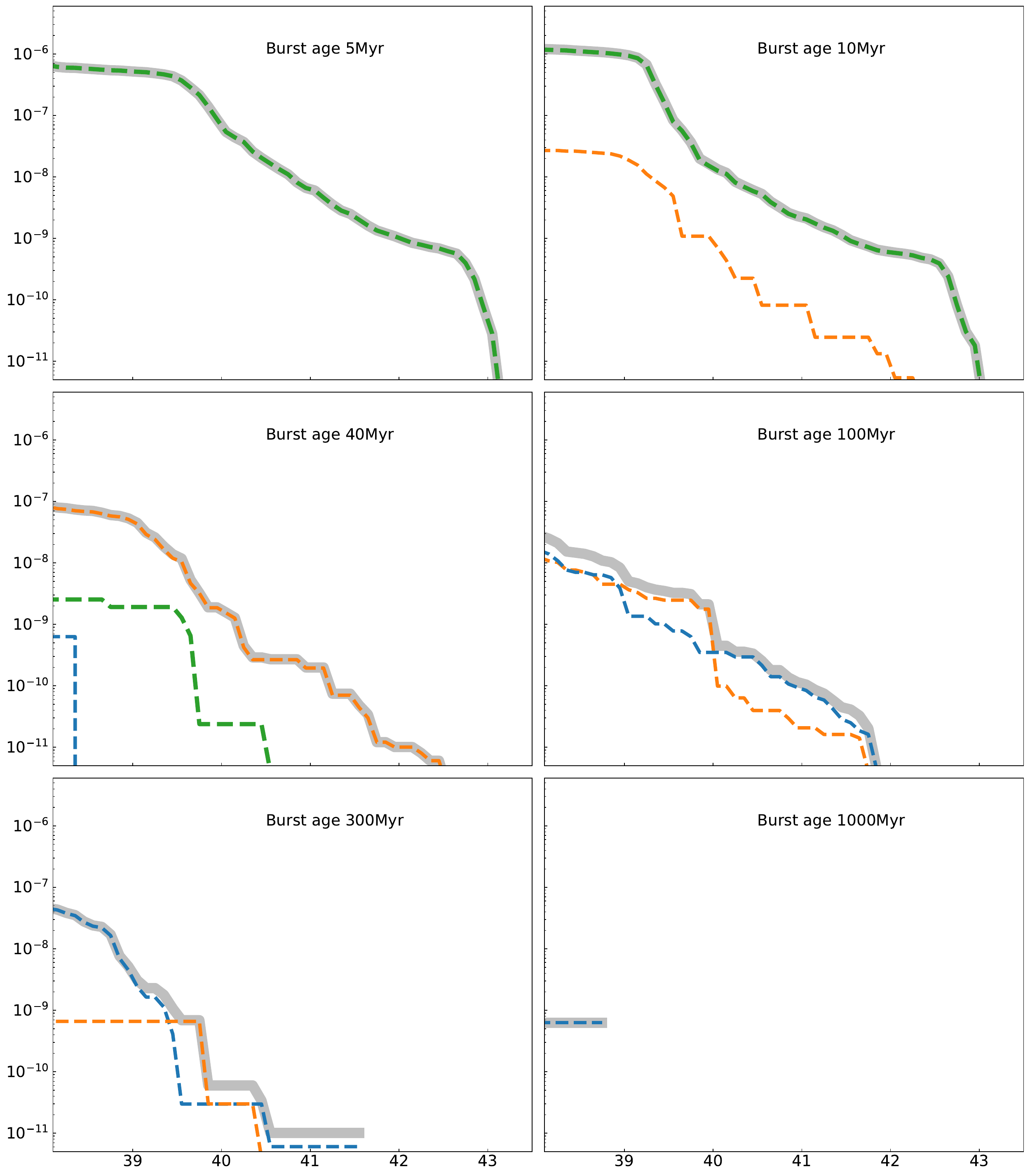}};
\node[node distance=0cm, yshift=-10.5cm] {log$_{10} (L_{\rm X}/\mathrm{erg\,s^{-1}}$)};
\node[node distance=0cm, rotate=90, anchor=center,yshift=9.0cm] {N$(>L_{\rm X})$ / $\mathrm{M_\odot}$};
\end{tikzpicture}
\caption{Synthetic XLFs of burst populations for model\,A. The panels show the different types of XRBs, namely, HMXBs, IMXBs, and LMXBs, as described in the legend. The combined XLF is shown by the dotted-gray line.}
\label{fig:combined_type_m64}
\end{figure*}

\section{Results}
\label{sec:results}

Stars of various masses evolve differently with time, leading to different demographics of the ULX populations from progenitors formed at different ages. Using population synthesis, we investigated the effect of age on ULXs. Figure\,\ref{fig:xrb_number} shows the distributions of isotropic-equivalent X-ray luminosities (calculated using the description in Section\,\ref{sec:methods:x-ray_lum}) of ULXs in the burst populations (for both the models\,A and B) without down-weighting the geometrically beamed ULXs. At high luminosities in an observed population, the sources would be beamed, and only a fraction of them would be observed. Overall, as the burst age increases, the number of ULXs decreases. This is because older populations are dominated by LMXBs that are comparatively less likely to reach ULX-like luminosities than HMXBs. For both models, the oldest population (with the burst age of $1000\,\rm Myr$) has no ULXs and, therefore, does not appear in the figures presented. The highest luminosity reached for XRBs with age 1000\,Myr for model\,A is $6\times 10^{38}\, \rm erg\, s^{-1}$, while for model\,B, it is $5\times 10^{37}\, \rm erg\, s^{-1}$ (well below our X-ray threshold for ULXs).

Going from model\,A to B, the number of ULXs reduces overall, while the peak luminosities remain similar (see Figure\,\ref{fig:xrb_number}). The effect is stronger for younger populations ($N_{\rm model\, A}/N_{\rm model\, B}\approx 3, 4, 7 $, corresponding to burst ages of 5\,Myr, 10\,Myr, and 40\,Myr, respectively), while older populations do not show the same trend ($N_{\rm model\, A}/N_{\rm model\, B}\approx 0.7$). Younger populations are dominated by BH accretors, which are affected more by the kick normalization assumed with respect to older populations. For model\,A, BH kicks are normalized using the mass of the newly formed BH, causing heavy BHs to receive negligible kicks, while the kicks for model\,B are not normalized and receive NS-like kicks. As seen in the parameter study by \citet{2023A&A...672A..99M}, different kick normalizations result in a reduction in the number of wind-fed XRBs with increasing kick strengths. The corresponding effect on RLO XRBs is not as straightforward. With increasing kick strength, though more binaries are disrupted, the number of SN surviving binaries that undergo RLO in close orbits increases since more post-SN binaries have high eccentricities. Therefore, even though we do see a slight reduction in ULXs with stronger kicks, there is not much change in the peak X-ray luminosity.  

Figure\,\ref{fig:combined_type_m64} shows the different burst populations sub-divided into the types of XRBs, specifically for model\,A (correspondingly, the burst populations for model\,B are shown in Fig. \ref{fig:combined_type_m44}), extending down to luminosities around $10^{38}\, \rm erg\, s^{-1}$, which is below our ULX threshold. As expected, younger populations are fully dominated by HMXBs (below 10\,Myr), and with age, the prevalence shifts to IMXBs (at ages 40 to 100\,Myr) and then to LMXBs (older than 100\,Myr). Populations older at 1000\,Myr do not contain ULXs. The dominance of IMXBs in ULXs lasts for a shorter duration (being prominent only at 40\,Myr) than LMXBs since IMXBs have shorter RLO phases. Their intermediate-mass donors cannot drive winds as high as those of high-mass donors until the RLO provides them with enough of a mass-transfer rate to emit X-rays. The behavior of the populations is similar between models\,A and B with respect to the burst ages. Therefore, IMXBs are viable candidates to explain ULX luminosities at these ages, reaching $\sim 10^{40}\, \rm erg\, s^{-1}$, as was also suggested by \citet{2017ApJ...846..170T} and \citet{2020A&A...642A.174M}.

\begin{figure}[!ht]
\centering
\includegraphics[width=\linewidth]{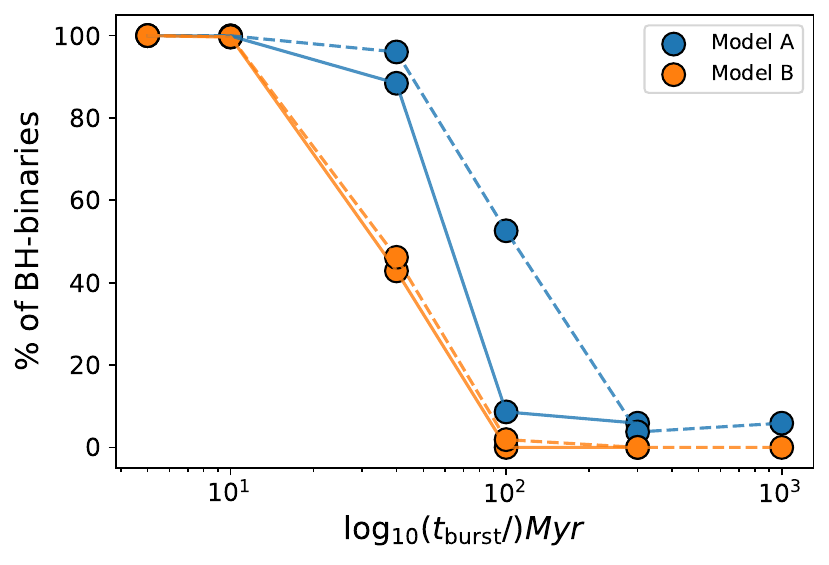}
\caption{Percentage of BH binaries present in the simulated ULXs (solid lines) and in the simulated XRBs (dashed lines) following the parameters of models\,A (shown by orange) and B (shown by blue).}
\label{fig:bh_total_v_age}
\end{figure}

\subsection{Nature of the accretors}\label{sec:results:acc_nat}

To investigate the nature of accretors present in ULXs, we looked at the population of accreting BHs with respect to the total XRB populations. Figure\,\ref{fig:bh_total_v_age} shows the percentage of BH XRBs (and BH ULXs) present in the total XRBs (and ULXs) across all the burst ages. The populations with the burst age of $1000\,\rm Myr$ (for both models\,A and B) have no ULXs. Therefore, they are not included in this figure. At 5\,Myr, the populations have no NSs since the population is too young for NS progenitors to evolve out of the main sequence (MS), and they are 100\% dominated by accreting BHs. The stars that would form NSs have initial masses in the approximate range of 8 to 25\,M$_{\odot}$, with the IMF favoring the lower masses. The MS lifetime for the stars within this range is $\sim 1$ to 10\,Myr. Since lower masses dominate the populations, many of the NS progenitor stars are still in their MS. The number of BH XRBs (and BH ULXs) decreases with time as NSs increase in number because the donors present in BH XRBs are mostly in the intermediate- to high-mass range. With time, an increasing number of these stars would undergo core collapse. Due to stronger BH kicks in model\,B, the number of BH binaries surviving the SN is less than in model\,A.

\begin{figure*}[htb]
\hspace{2.0cm}\includegraphics[width=0.7\textwidth]{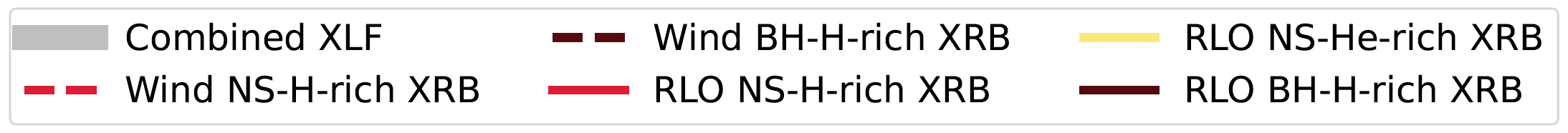}\\
\centering
\begin{tikzpicture}
\node (img1)  {\includegraphics[width=0.95\textwidth]{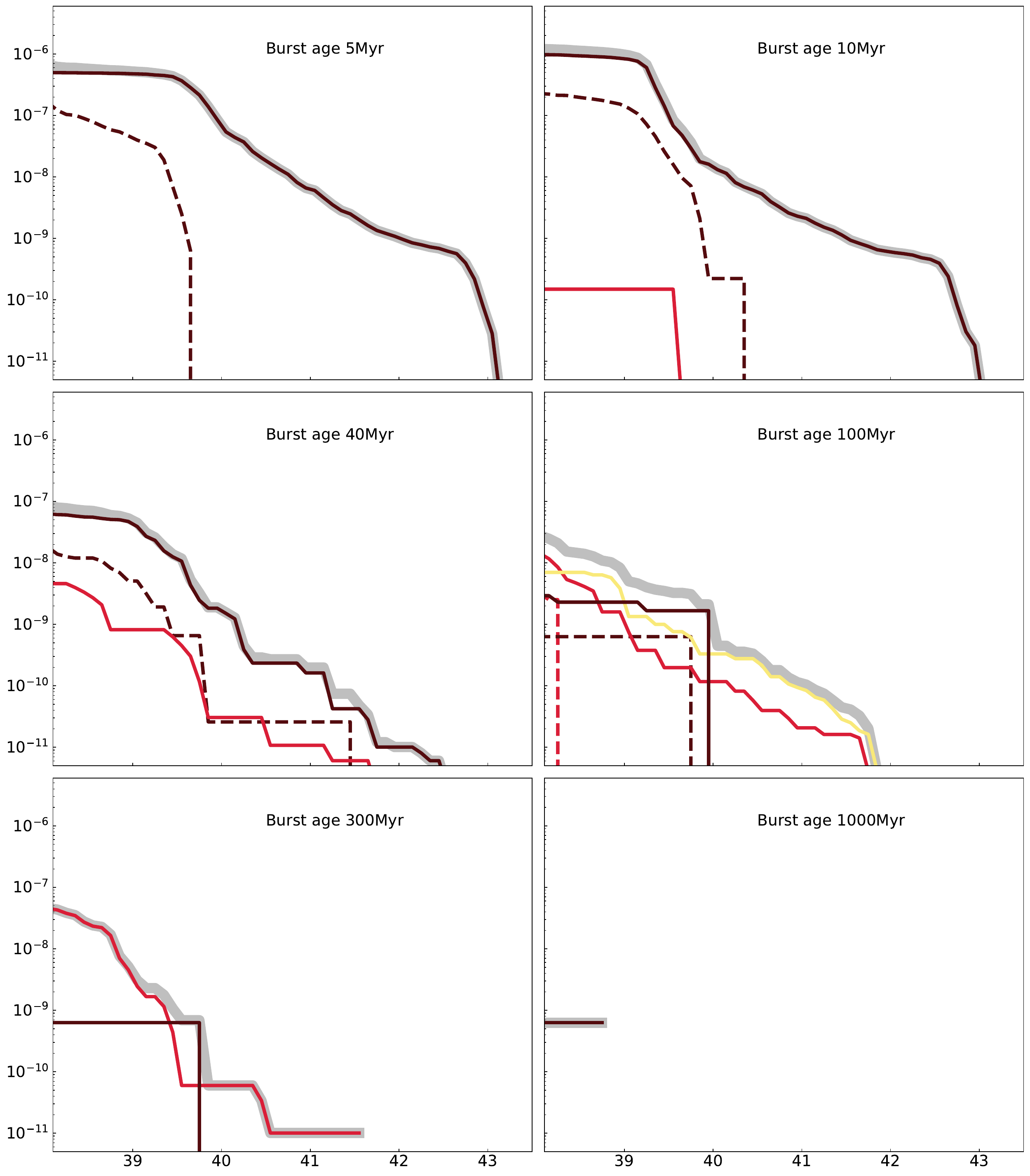}};
\node[node distance=0cm, yshift=-10.5cm] {log$_{10} (L_{\rm X}/\mathrm{erg\,s^{-1}}$)};
\node[node distance=0cm, rotate=90, anchor=center,yshift=9.0cm]  {N$(>L_{\rm X})$ / $\mathrm{M_\odot}$};
\end{tikzpicture}
\caption{Synthetic XLFs of burst populations for model\,A. The populations are further split into the type of the XRBs (RLO or wind), type of accretors (BH or NS) and type of donors (H-rich or He-rich). The combined XLF is shown by the dotted-gray line.}
\label{fig:combined_xlf_m64}
\end{figure*}

\begin{figure*}[htb]
\begin{minipage}{.5\textwidth}
  \includegraphics[width=.9\textwidth]{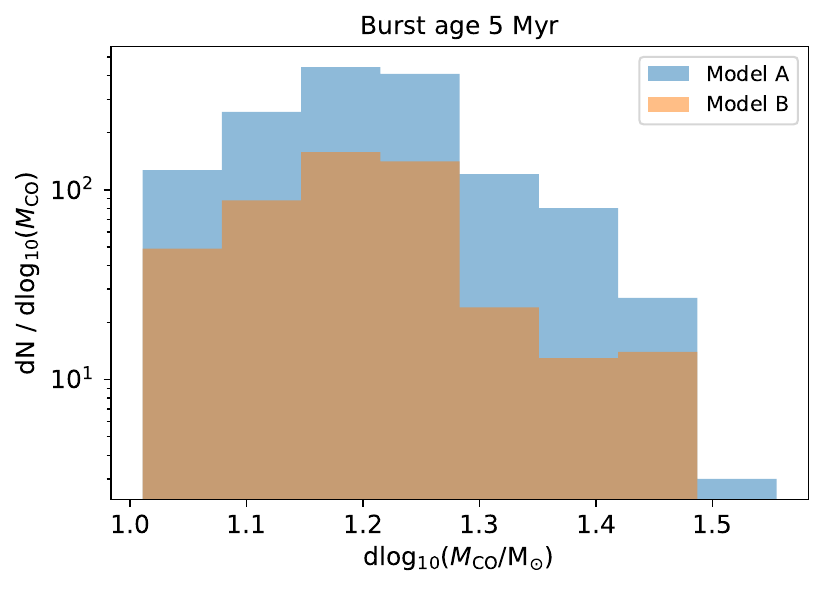}
\end{minipage}%
\begin{minipage}{.5\textwidth}
  \includegraphics[width=.9\textwidth]{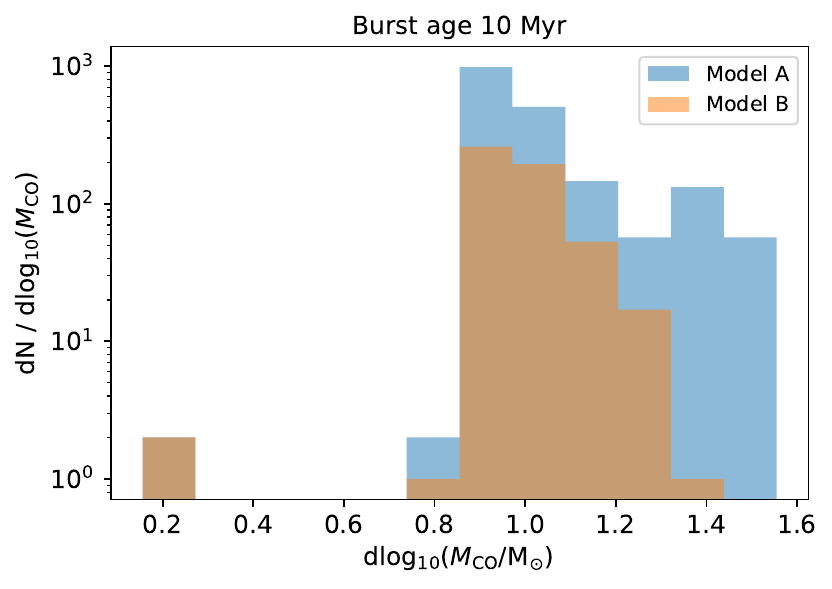}
\end{minipage}
\begin{minipage}{.5\textwidth}
  \includegraphics[width=.9\textwidth]{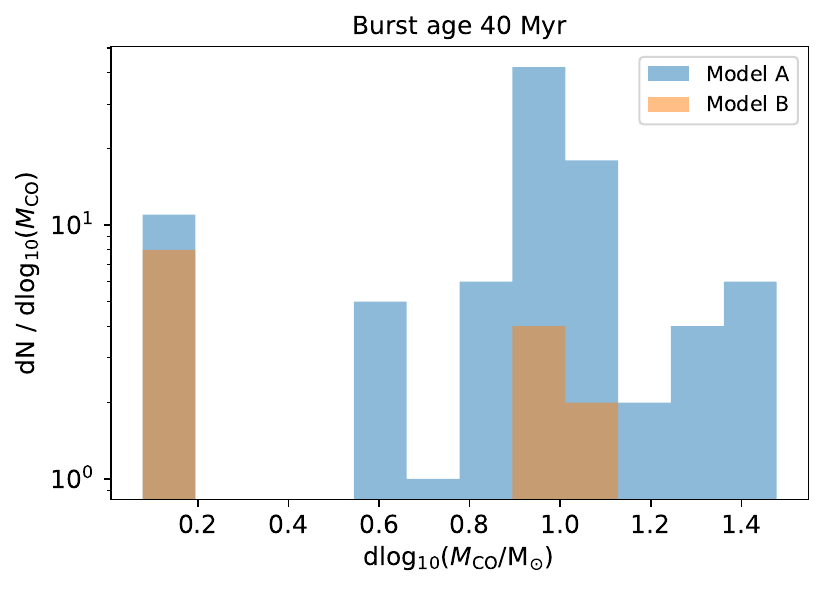}
\end{minipage}%
\begin{minipage}{.5\textwidth}
  \includegraphics[width=.9\textwidth]{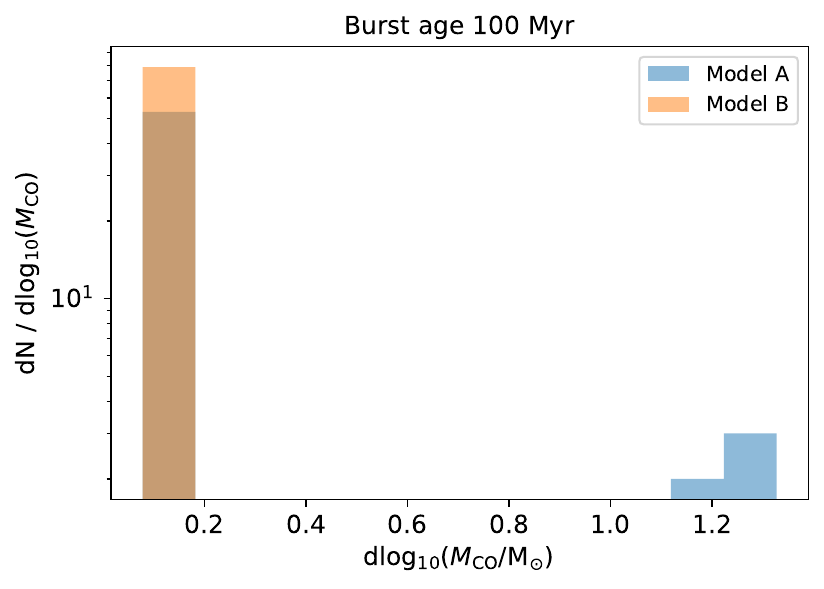}
\end{minipage}
\begin{minipage}{.5\textwidth}
  \includegraphics[width=.9\textwidth]{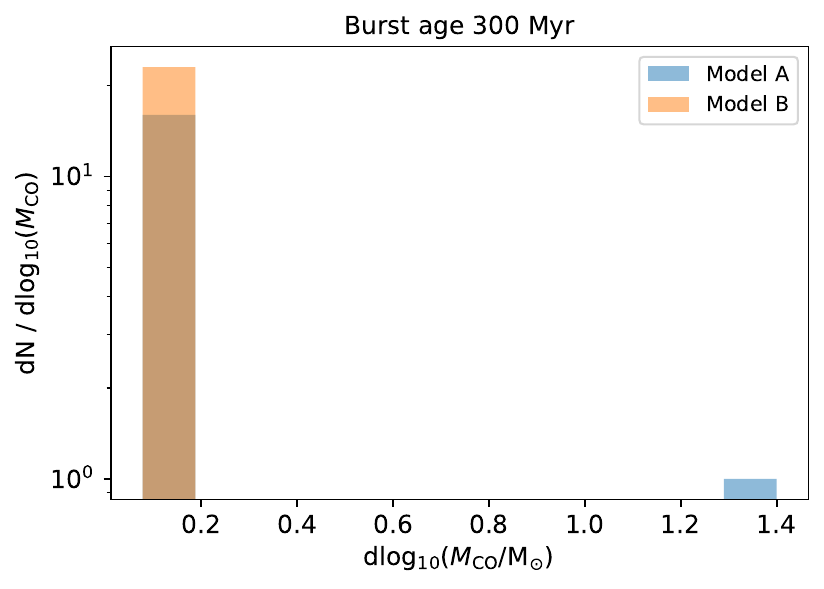}
\end{minipage}%
\begin{minipage}{.5\textwidth}
  \caption{Distribution of accretor masses for ULXs in the synthetic populations across ages 5 to 300\,Myr.}
  \label{fig:acc_dist}
\end{minipage}%
\end{figure*}

\begin{figure*}[htb]
\begin{minipage}{.5\textwidth}
  \includegraphics[width=.9\textwidth]{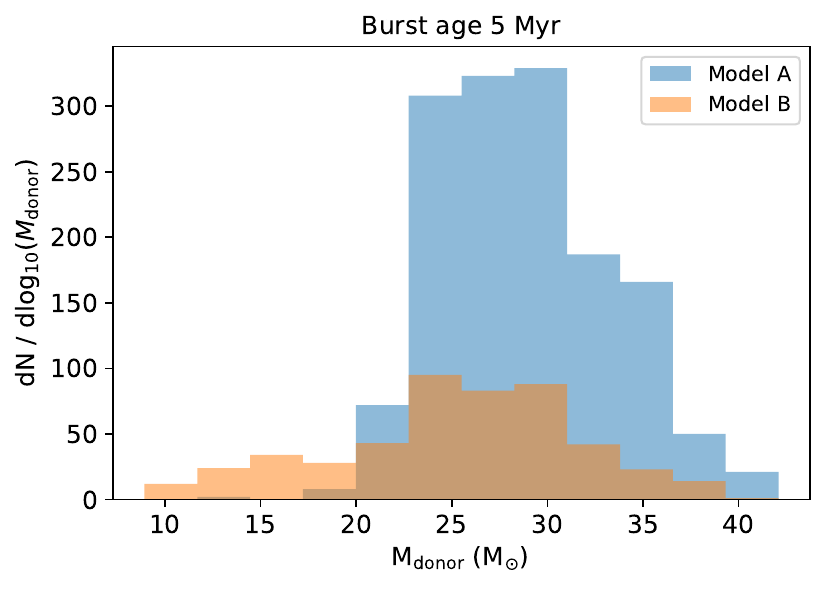}
\end{minipage}%
\begin{minipage}{.5\textwidth}
  \includegraphics[width=.9\textwidth]{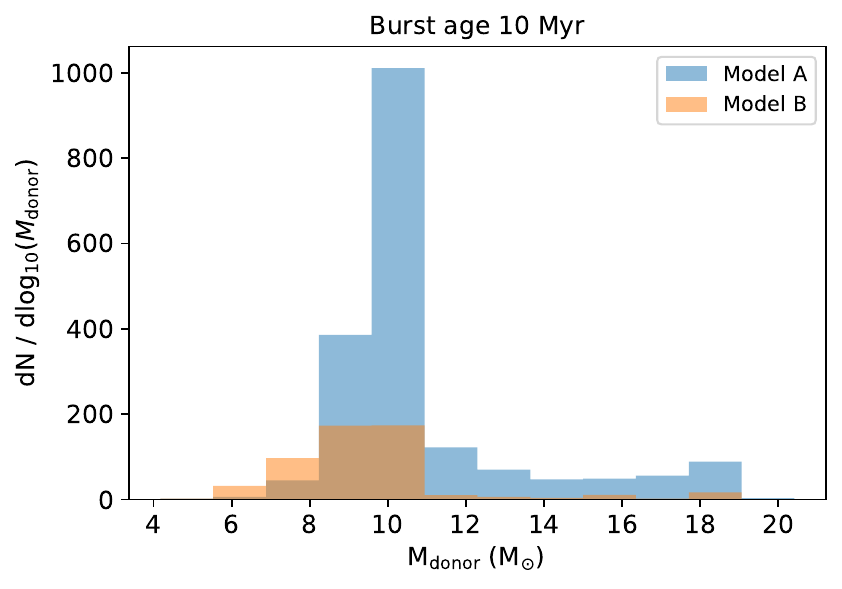}
\end{minipage}
\begin{minipage}{.5\textwidth}
  \includegraphics[width=.9\textwidth]{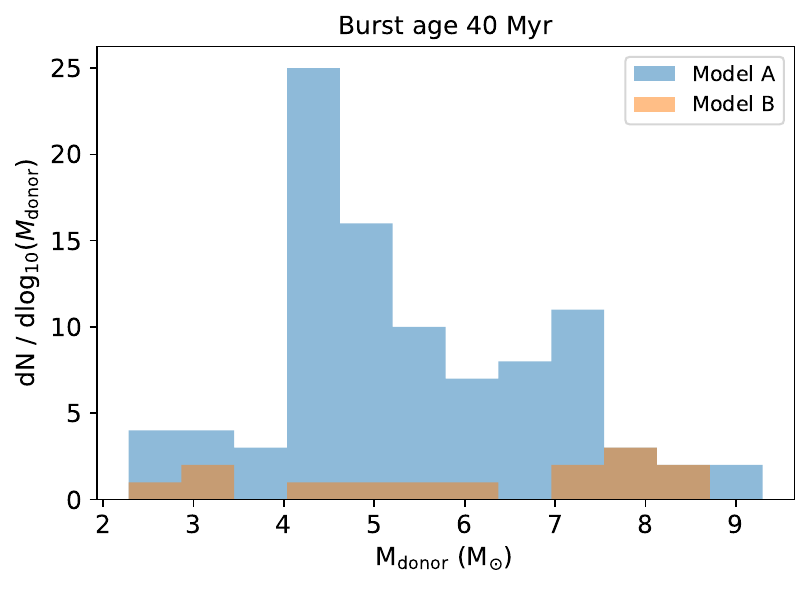}
\end{minipage}%
\begin{minipage}{.5\textwidth}
  \includegraphics[width=.9\textwidth]{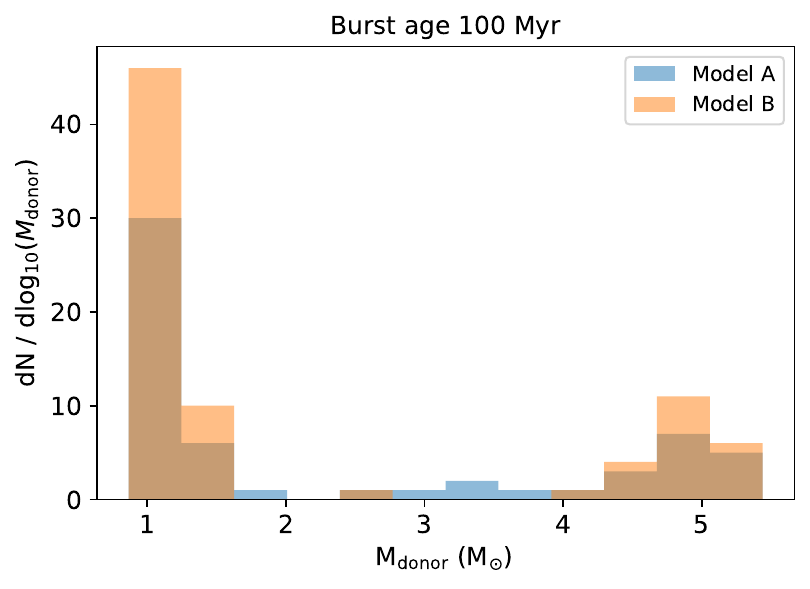}
\end{minipage}
\begin{minipage}{.5\textwidth}
  \includegraphics[width=.9\textwidth]{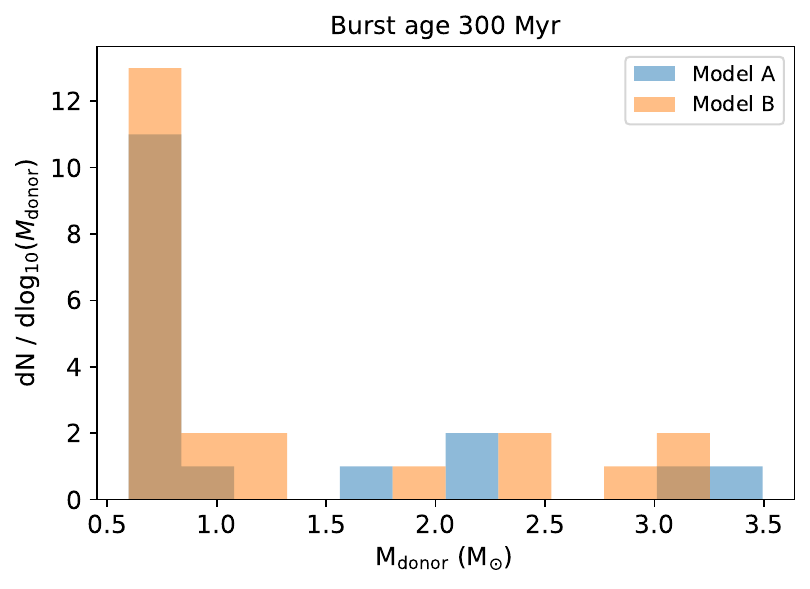}
\end{minipage}%
\begin{minipage}{.5\textwidth}
  \caption{Distribution of donor masses for ULXs in the synthetic populations across ages 5 to 300\,Myr.}
  \label{fig:don_dist}
\end{minipage}%
\end{figure*}

\begin{figure*}[htb]
\begin{minipage}{.5\textwidth}
  \includegraphics[width=.9\textwidth]{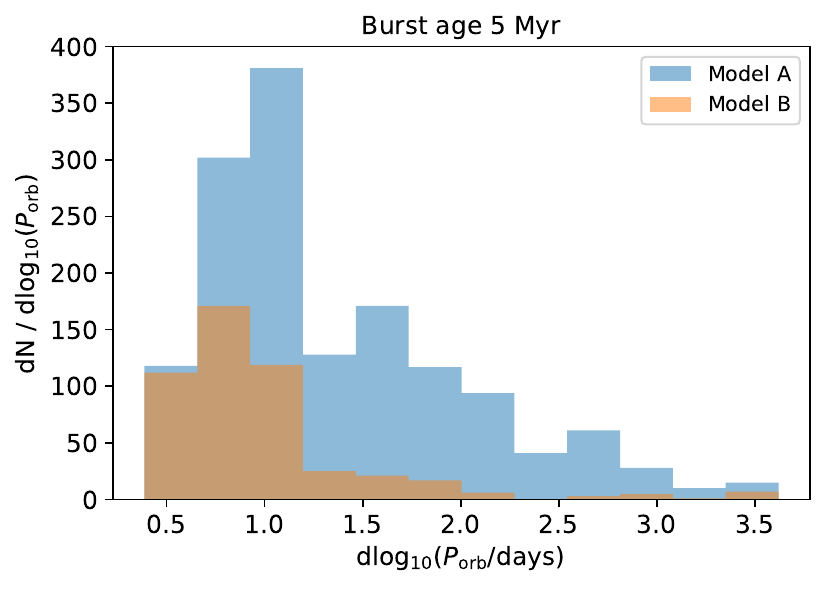}
\end{minipage}%
\begin{minipage}{.5\textwidth}
  \includegraphics[width=.9\textwidth]{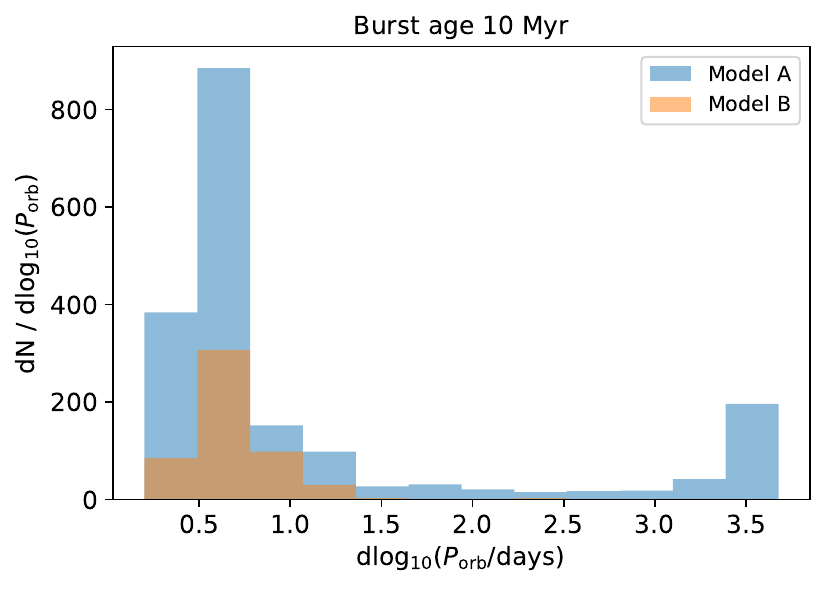}
\end{minipage}
\begin{minipage}{.5\textwidth}
  \includegraphics[width=.9\textwidth]{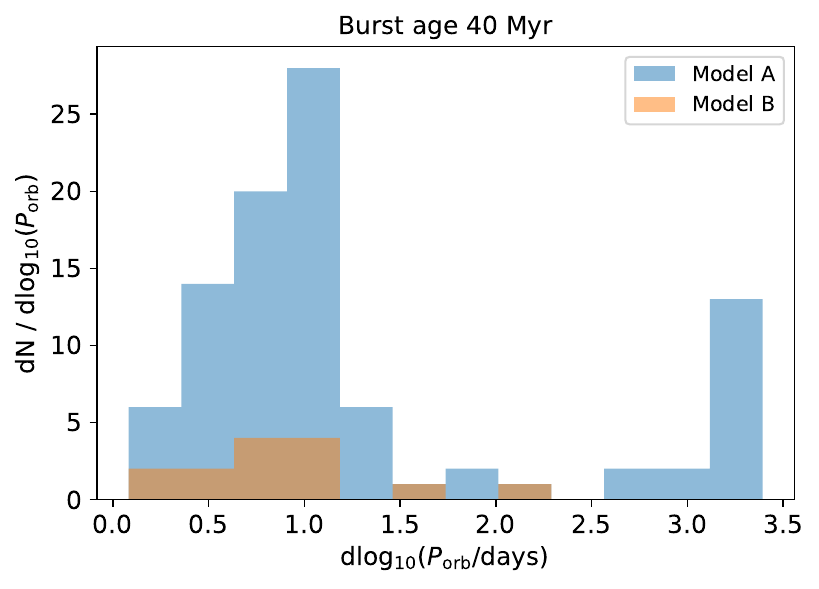}
\end{minipage}%
\begin{minipage}{.5\textwidth}
  \includegraphics[width=.9\textwidth]{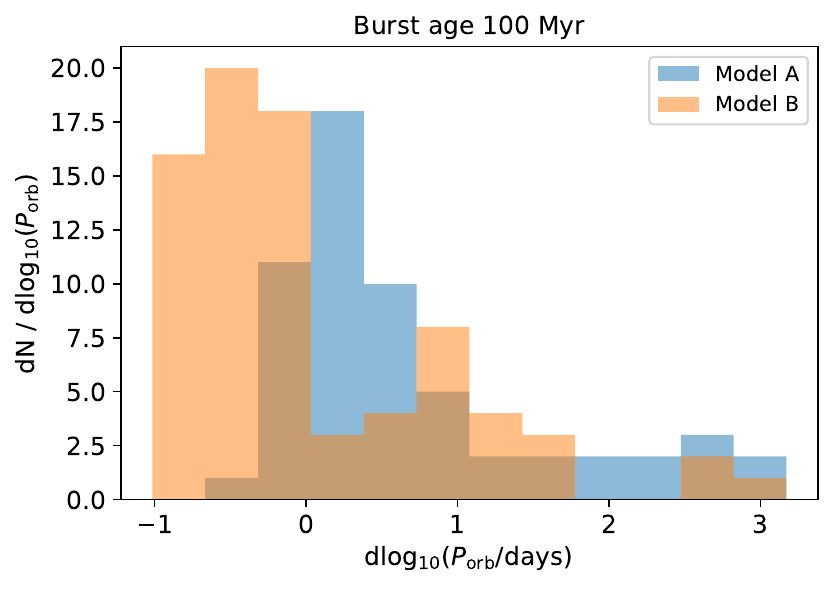}
\end{minipage}
\begin{minipage}{.5\textwidth}
  \includegraphics[width=.9\textwidth]{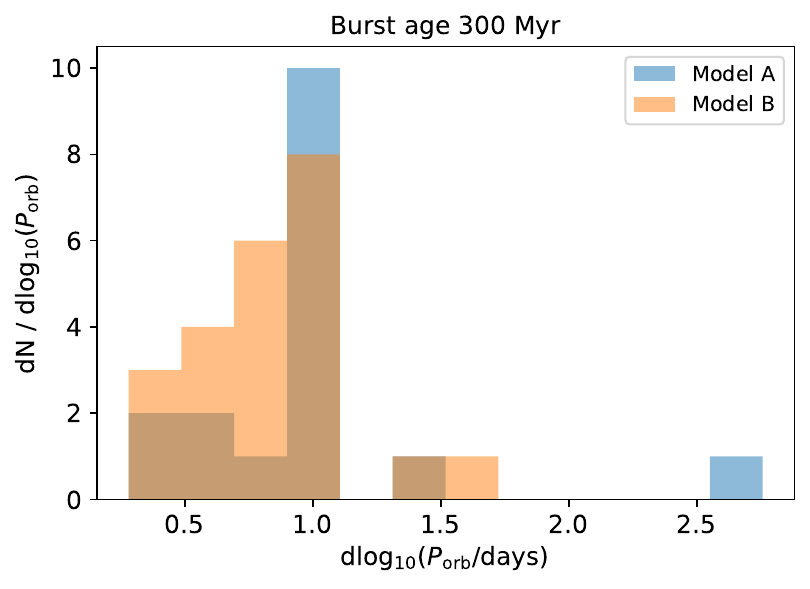}
\end{minipage}%
\begin{minipage}{.5\textwidth}
  \caption{Distribution of orbital periods for ULXs in the synthetic populations across ages 5 to 300\,Myr.}
  \label{fig:orb_dist}
\end{minipage}%
\end{figure*}

\subsection{Properties of typical ultra-luminous X-ray sources in each population} 

To get a better idea of the properties of ULXs in the synthetic burst populations, we looked at the synthetic XLFs of the populations. Figure\,\ref{fig:combined_xlf_m64} shows the sub-populations of each of the simulated populations, describing the type of the XRB (RLO or wind), type of accretor (BH or NS), and type of donor (H-rich or He-rich) for model\,A. The same for model\,B is shown in Figure\,\ref{fig:combined_xlf_m44}. Again, the figures extend to luminosities of $\sim 10^{38}\, \rm erg\, s^{-1}$, which is below our ULX threshold. In general, all ULXs are dominated by XRBs undergoing RLO, with some contribution from wind-fed XRBs at wide orbits for model\,A. There are  significantly fewer wind-fed ULXs in the ULX populations when following model\,B due to a disruption of wide orbits from strong SN kicks (see Figure\,\ref{fig:combined_xlf_m44}). Additionally, the distributions of the orbital properties for all burst ages and both models are presented in Figures\,\ref{fig:acc_dist} (for accretor masses), \ref{fig:don_dist} (for donor masses), and \ref{fig:orb_dist} (for orbital periods). The orbital properties for the simulated ULXs are not significantly different between models B and A; hence, we primarily discuss the ULXs for model\,A and point out the differences with model\,B when significant.

% 5Myr 
\subsubsection{Burst age of 5~Myr}\label{sec:results:5Myr}
We utilized the aforementioned figures to derive the orbital properties of typical ULXs for all burst ages, described by the peaks of the respective distributions. For the population with a burst age of 5\,Myr, a typical ULX has a BH in the mass range of about 10 to 15\,M$_{\odot}$ (the minimum BH mass being 10.25\,M$_{\odot}$), with a massive 28\,M$_{\odot}$ H-rich MS donor undergoing RLO in an orbit of about 10\,days. The binary, after the primary SN event, evolves further until the massive donor fills its Roche lobe, at around 10\,days, and it starts an intense mass-transfer phase while it is still on the MS, leading to its appearance as a ULX. As the stronger kicks in model\,B are more prone to disrupting wider orbits, the number of ULXs drops off very quickly with an increasing orbital period compared to model\,A (see Figure\,\ref{fig:orb_dist}). The distribution of the donor masses differs between the two models (see Figure\,\ref{fig:don_dist}). Model\,B has a flatter distribution compared to model\,A because the differing binaries between the two models are the ones that were disrupted by the strong kicks in model\,B. These binaries correspond to a wide range of orbital periods (many hundreds of binaries with $\gtrsim 3$\,days) and massive donor masses ($\gtrsim  25\,\rm M_{\odot}$). 

%and 10Myr
\subsubsection{Burst age of 10~Myr}
A population at 10\,Myr has a typical ULX, with a $9\,\rm M_{\odot}$ BH, a $10\,\rm M_{\odot}$ H-rich MS donor, and an orbital period of $4$\,days undergoing RLO. The distribution of the orbital parameters is similar to the ULXs at 5\,Myr (see Section \ref{sec:results:5Myr}), although with the difference of a shift in the donor mass to less massive (see Figure\,\ref{fig:don_dist}) and more compact stars, and hence the difference also corresponds to narrower orbits, as can be seen in Figure\,\ref{fig:orb_dist}. The radii of donors at 5\,Myr peaks at around 36\,R$_{\odot}$, while for 10\,Myr the peak is at 12\,R$_{\odot}$. Additionally, there is an emergence of NS accretors at this age.

% 40 Myr
\subsubsection{Burst age of 40~Myr}
At 40\,Myr, a typical ULX has a similar CO mass as that at 10\,Myr, with a 9\,M$_{\odot}$ accreting BH and an intermediate-mass (around $4.5\,\rm M_{\odot}$) H-rich MS donor in an RLO orbit with a period of $10$\,days. The peak of the orbital period distribution is larger than that at 10\,Myr because at this age (see Figure\,\ref{fig:orb_dist}), even though the typical donor star is less massive than before, the mass is being transferred from a lower-mass star to a more massive CO, leading to orbital expansion, which contrasts with the process at the younger age when the mass ratios were more even. There is another peak in the ULX population at orbital periods $\gtrsim 1000$\,days (only for model\,A) corresponding to H-rich post-MS donors that have expanded to large radii and are transferring matter effectively via wind-fed accretion. This population of wind-fed BH XRBs can already be seen in model\,A at 10\,Myr. However, as can be seen when comparing Figures\,\ref{fig:combined_type_m64} and \ref{fig:combined_type_m44}, the wind-fed HMXBs disappear completely in the population following model\,B due to the strong BH kicks disrupting wide orbits that would have led to wind-fed accretion. Additionally, NS accretors begin to gain noticeable numbers at this age (see Figure\,\ref{fig:acc_dist}), with about 5\%\ and 40\% of NSs present in the XRBs (for models\,A and B, respectively), as can be seen in Figure\,\ref{fig:bh_total_v_age}. 

% 100 Myr
\subsubsection{Burst age of 100~Myr}
At 100\,Myr, a typical ULX has an NS accretor with a mass of $1.2\,\rm M_{\odot}$ and a low-mass stripped-helium (stripped-He) donor with $1.2\,\rm M_{\odot}$ in an orbit of 2\,days. After the primary SN event, the donors are H-rich MS stars and are in wide orbits (centered around 1000\,days). Consequently, when the donors fill their Roche lobes, they have depleted their core-He. Since they are rapidly expanding He-giants, the binary interaction is more prone to instability, and a CE phase occurs that strips the outer H-rich envelope of the donor. The surviving NS+He stripped binary then evolves until the donor fills its Roche lobe again, starting its XRB phase. The peak of the orbital period distribution for model\,B is at a much smaller orbit, at $0.5$\,days, compared to model\,A (which is at 2\,days), as shown in Figure\,\ref{fig:orb_dist}. Since the effect of different BH natal kicks is not applicable to a majority of the population at this burst age, the secondary difference between the models becomes more apparent, such as CE efficiency or $\alpha_{\rm CE}$. The value of the $\alpha_{\rm CE}$ is approximately proportional to the final orbital separation at the end of the CE phase and determines whether this interaction is stable or not. Going from model\,A to model\,B, the $\alpha_{\rm CE}$ decreases from 1.0 to 0.3, which would intuitively lead to more binaries with narrower post-CE orbits and higher instability \citep[refer to ][for a detailed study on the effect of various assumptions of the CE phase on XRBs]{2023A&A...672A..99M}. However, in the context of ULXs, the model\,B ULXs experience an increase in the fraction of binaries with narrower orbits after the CE in the surviving binaries. Hence, the difference in the distributions of orbital periods in Figure\,\ref{fig:orb_dist}.

In the distribution of the ULX donor masses shown in Figure\,\ref{fig:don_dist}, apart from the He-rich donors, there is another population of intermediate-mass stars (around 5\,M$_{\odot}$). The same binaries result in the second peak in the orbital periods seen for model\,B in Figure\,\ref{fig:orb_dist}. These ULXs correspond to post-MS donors that previously underwent stable RLO and did not lose their H envelopes. For these binaries (and for both sets of model assumptions), RLO begins after the first SN when the donor is an H-giant star (in the H shell-burning phase). The relative expansion of the donor radius with respect to the Roche-lobe radius is not as high as when the donors are He-giants. The stability of the mass transfer is linked to the evolutionary state of the donor \citep{2020A&A...642A.174M}, as stars at certain later evolutionary stages have large radii with deep convective regions that expand rapidly on mass loss. If the rapid mass loss exposes an underneath radiative layer, the radius contracts, stabilizing the binary. Otherwise, rapidly expanding donors would lead to a CE.

\subsubsection{Burst age of 300~Myr}
% 300 Myr
A typical ULX at 300\,Myr has an NS accretor with $1.2\,\rm M_{\odot}$ and a low-mass post-MS H-rich donor with $0.6\,\rm M_{\odot}$ in an orbit of $8$\,days. For model\,B, there is an additional XRB population with stripped-He donors that is not present in model\,A. However, their luminosities do not cross our ULX threshold of $10^{39}\,\rm erg\, s^{-1}$. These XRBs dominate the population at luminosities $\lesssim 10^{38}\, \rm erg\, s^{-1}$ at this age while appearing as ULXs at 100\,Myr.

% 1000 Myr: removed since no ULXs
%In the small XRB population created from the 1000\,Myr population, most binaries are RLO NS XRBs with H-rich donors (with the majority of donors $\lesssim 2.0\,\rm M_{\odot}$), with X-ray luminosities below $10^{37}\, \rm erg\, s^{-1}$. NSs are expected to dominate old stellar populations, and stars with initially low masses at ZAMS have a long enough MS lifetime to initiate RLO at these ages. For model\,A, however, there is one RLO BH XRB, having a $21.4\,\rm M_{\odot}$ BH and a $1.9\,\rm M_{\odot}$ H-rich post-MS donor, in a $56$\,day orbit, while model\,B has no XRBs with BH accretors. 

\begin{figure}[!ht]
\centering
\includegraphics[width=\linewidth]{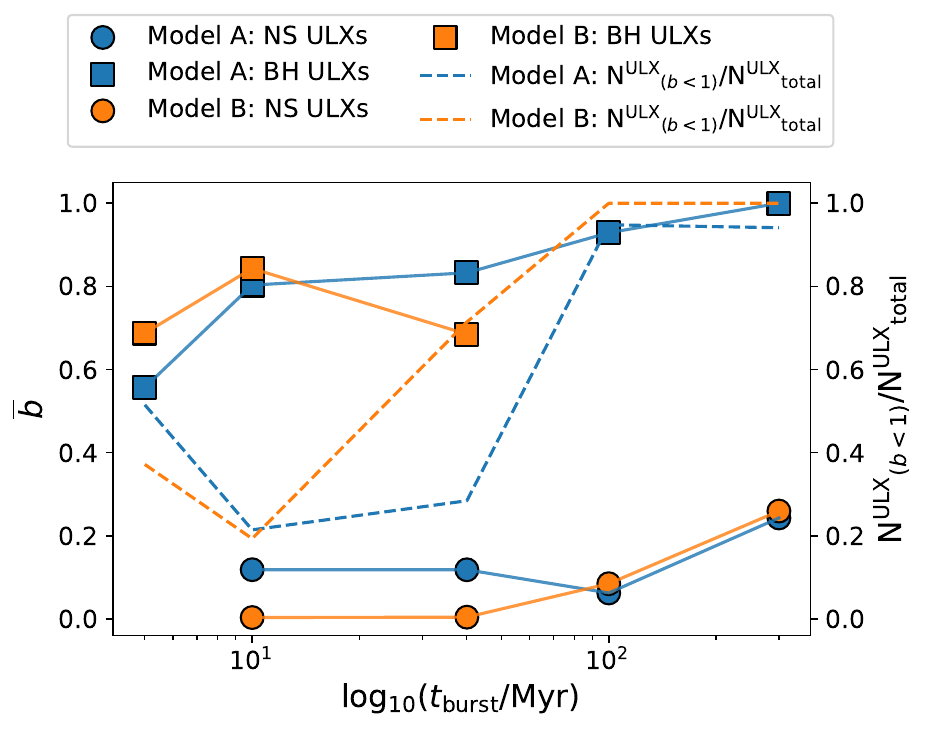}
\caption{Average beaming factor $(\,\bar{b}\,)$ for the ULX populations spanning all burst ages, distinguishing between the type of accretor (NS, shown by circles, or BH, shown by squares). Populations that did not have any ULXs are not shown. The figure also shows the fraction of beamed (with $b<1.0$) to total ULXs in each population, shown as dashed lines. The blue colored symbols and lines correspond to the results for model\,A, and the orange-colored symbols and lines are for model\,B.}
\label{fig:mean_beam_v_age}
\end{figure}

\subsection{Beamed X-ray emission} 

An aspect that would affect the observed versus underlying population of ULXs is the effect of geometrical beaming. The stronger the geometrical beaming of the X-ray emission, the smaller is the chance of observing the ULX. In our synthetic populations, we defined beamed emission using the factor $b$, which is defined in Equation\,\ref{eq:c4_beaming}, with beamed emission as $b<1$ and unbeamed emission as $b=1$. Hence, the observed luminosities from super-Eddington accretion disks are enhanced by the factor $b$.

Figure\,\ref{fig:mean_beam_v_age} shows the average beaming factor for each of the ULX populations, with the populations categorized by the type of accretor (NS or BH). We note that the populations with no ULXs (at 1000\,Myr), no NS ULXs (at 5\,Myr), or no BH ULXs (at 100 and 300\,Myr for model\,B) are not shown. The figure also shows the fraction of beamed ULXs in each population. As expected, NSs are more strongly beamed than BHs, with their beaming factors less than approximately 0.25 across all burst ages and both models. The BHs have relatively less beamed emission and are visible more than 50\% of the time whenever they are present in the population. With age, the average beaming factor for BH ULXs increases, that is, their emission becomes more isotropic on average, in agreement with \citet{2019ApJ...875...53W}. This effect is stronger for model\,A, unlike model\,B where BHs disappear from the population by 100\,Myr. For the populations with no NS ULXs at 5\,Myr, the beamed-to-unbeamed ratio is almost equal, with a slight dominance for unbeamed ULXs in model\,B, as strong kicks in model\,B reduce the number of close BH binaries that would undergo RLO and have high mass-transfer rates. 

At 10\,Myr, the beamed sources decrease to about 20\% for both models while still being dominated by BH ULXs. The reason for the decrease in the number of beamed ULXs is the shift to less massive donor masses as the population age increases (the peak of the donor mass distribution goes from $\sim 25\,\rm M_{\odot}$ to $\sim 10\,\rm M_{\odot}$; see Figure\,\ref{fig:don_dist}). These less massive donors have longer MS lifetimes and smaller radii compared to the massive donors at 5\,Myr. Therefore, a smaller fraction of donors has super-Eddington mass-loss rates that require considerable overflowing of the Roche lobe during RLO, resulting in decreased beaming \citep[for mass transfer rates below 8.5 times the Eddington limit, the beaming factor is one, following the model from][]{2009MNRAS.393L..41K}. Less massive MS donors typically do not drive mass-transfer rates to values as high as more massive MS donors. This is also the reason for the decrease in geometrical beaming for BH ULXs at 10\,Myr. The small population of NS ULXs that appears at this burst age is fully beamed. 

For model\,A at the burst age of 40\,Myr, the ratio of beamed versus unbeamed ULXs remains similar to the one at 10\,Myr, but the ratio is reversed for model\,B, as most ULXs become beamed ($\sim 71\%$). The reason for this discrepancy is the change in the demographics of the beamed ULXs between the two models. Due to the difference in BH kicks, the ULX population for model\,B is dominated by NSs, which are always beamed, owing to their Eddington limits that are an order of magnitude lower than BH XRBs. For ages greater than $40\,\rm Myr$, the fraction of beamed ULXs is around $\gtrsim 94\%$ in model\,A and $100\%$ in model\,B since both populations are dominated by NS ULXs. As can be seen in Figure\,\ref{fig:mean_beam_v_age}, NS ULXs, whenever they are present, are strongly beamed, while BH ULXs are not as beamed across all the burst ages.

\begin{figure}[!ht]
\centering
\includegraphics[width=\linewidth]{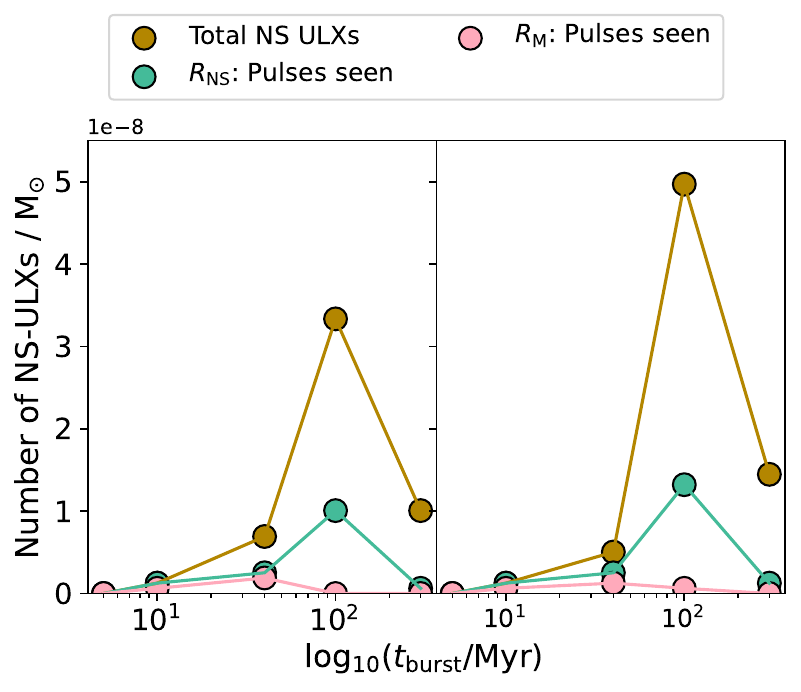}
\caption{The number of NS ULXs for models\,A and B (left and right panels, respectively) per stellar mass in the burst populations. With green filled circles, we show the number of NS ULXs with observable pulses using the NS radius ($R_{\rm NS}$) as the accretion radius. With pink circles we show the number of NS ULXs with observable pulses using the magnetospheric radius ($R_{\rm M}$), and the mustard circles show the total number of NS ULXs.
}
\label{fig:pulses}
\end{figure}

\subsection{Suppression of pulses caused by accretion onto a neutron star}
\label{sec:results:sup_pulse}

It has been suggested that the accretion phase would affect the observations of pulses seen from an NS accretor \citep{2017MNRAS.467.1202M,2020MNRAS.494.3611K}. Using Equations\,\ref{eq:delta_m_Rns} and \ref{eq:delta_m_Rm} to define the accreted matter (at various radii) required to align the NS spin axis to the orbital axis, we modeled the suppression of pulses and investigated the relative contribution of pulsed versus non-pulsed systems in ULX populations as well as the distribution of the types of COs (BHs and NSs). In our investigation, the term "non-pulsed" simply refers to NSs that have accreted matter greater than the amounts described by Equations\,\ref{eq:delta_m_Rns} and \ref{eq:delta_m_Rm}, as their X-ray pulsations are suppressed. Figure\,\ref{fig:pulses} shows the number of NS ULXs per stellar mass for models\,A and B, depicting the total population and the NS ULXs that would have observable pulses for angular momentum accreted at two radii. The two radii in question are the NS radius ($R_{\rm NS}$) and the magnetospheric radius ($R_{\rm M}$). The populations that are 10\,Myr old and have a very small number of NS ULXs present ($\sim 10^{-9}/\,\rm M_{\odot}$) all have observable pulses. However, since the populations at 10\,Myr are dominated by BH ULXs (almost 100\%), the chance to observe these NS ULXs would be insignificant. 

For other ages, we observed that a small fraction of the total NS ULXs emit observable pulses, especially for populations with ages from 40\,Myr to 100\,Myr (when NSs start to dominate the ULX population) with pulses from NS ULXs observable for 30\%\ to 50\% of the binaries. An important feature is present at 100\,Myr, where the pulses from about $30\%$ of NS ULXs may be observed when the accretion occurs at the $R_{\rm NS}$ and $\lesssim 2\%$ when accretion is at $R_{\rm M}$. The matter accreted at $R_{\rm M}$ carries more angular momentum compared to $R_{\rm NS}$ due to its larger distance. Therefore, less material needs to be accreted for the pulses to be suppressed, and the number of NS ULXs showing pulses is significantly lower for accretion at $R_{\rm M}$. For a typical NS (with surface magnetic field $\sim 10^{12}\,\rm G$), the magnetospheric radius is around $2\times 10^{8}\rm\,cm$. However, for an NS with higher field strength, such as $\sim 10^{14}\,\rm G$ (i.e., a magnetar), the magnetospheric radius is larger by a factor of around $14$. This increased accretion radius would lead to a faster suppression of observable pulses, as the angular momentum transferred from the accreted material would be larger than the accretion from a smaller radius. Consequently, magnetar ULXs with pulses are more difficult to observe. At 300\,Myr, the fraction of NS ULXs with observed pulses drops down to 5\% for accretion at $R_{\rm NS}$ and to 0\% for accretion at $R_{\rm M}$ since the accreted matter increases with time, eventually crossing the threshold set by Equations\,\ref{eq:delta_m_Rns} and \ref{eq:delta_m_Rm}.

Using Equations\,\ref{eq:delta_m_Rns} and \ref{eq:delta_m_Rm}, we estimated the timescales of the accretion phases in the ULXs. Varying the NS spin period between 0.1 and 10\,s, the mass between 1.2 and 2.0\,M$_{\odot}$, and the magnetic field strength between $10^{8}$ and $10^{14}\rm\,G$, the resulting timescales are between 1 and $10^5$\,yrs (over both estimates of the accretion radius). The medians for the timescale distributions are around 500\,yrs (for accretion at $R_{\rm NS}$) and around 100\,yrs (for accretion at $R_{\rm M}$). While these periods of time imply that most NS ULXs should not be seen, Figure\,\ref{fig:pulses} clearly shows about 30\% of the observed NS ULXs at 100\,Myr (for model\,A). This unexpectedly high number can be explained by the fact that most of the NS ULXs (between the ages 40 and 100\,Myr) that have any observable pulses have H-giant donors, whereas the rest of the hidden population is dominated by either stripped-He or H-rich MS donors. Stable mass transfer for giants donor stars occurs over thermal timescales, and for MS donors stars, it occurs over nuclear timescales (a few orders of magnitude longer than the thermal timescales). Hence, even though the mass-transfer rates of the mass transfer from giant stars (with deep convective envelopes) go to high values (often exceeding the Eddington limit of the accretor), the phase itself is short lived. At 10\,Myr, the few NS ULXs present have pulsed emission, as the systems are young and not enough time has passed to fully align the spin and beaming axes. At 300\,Myr, both sets of NS ULXs (pulsed and non-pulsed) have H-giant donors, with higher masses ($\sim 3\rm\,M_{\odot}$) tending toward the pulsed NS ULXs and lower masses ($\lesssim 2\rm\,M_{\odot}$) going toward non-pulsed.

% ALL NS ULXs
% 10Myr, pulsed HMS donors no non-pulsed
% 40Myr, pulsed H-giant, non-pulsed HMS
% 100Myr, pulsed H-giant, non-pulsed He-MS
% 300Myr, pulsed H-giant, non pulsed H-giant

Comparing the NS ULX populations between the two models in Figure\,\ref{fig:pulses}, we found an increase in the total number of NS ULXs when going from model\,A to model\,B, most notably at the burst age 100\,Myr, where the total NS ULX number increases by about 1.5 times. The reason is the difference in the values of the $\alpha_{\rm CE}$ between the models (see Table\,\ref{table:c4_best_fit}). Lower values of $\alpha_{\rm CE}$ lead to narrower orbits of binaries that survive the CE phase. These narrow binaries lead to higher mass-transfer rates during RLO and therefore a higher number of ULXs. However, this increase in the number of NS ULXs with decreasing $\alpha_{\rm CE}$ was not as distinctly reflected in the number of ULXs where escaped pulses could be seen. This occurs because even though more NS binaries are in the ULX phase, due to a high degree of super-Eddington mass transfer, more of them are able to accrete enough material to fulfill the conditions described by Equations\,\ref{eq:delta_m_Rns} and \ref{eq:delta_m_Rm}. Hence, the number of NS ULXs with observable pulses increases only marginally.

\begin{figure}[!ht]
\centering
\includegraphics[width=\linewidth]{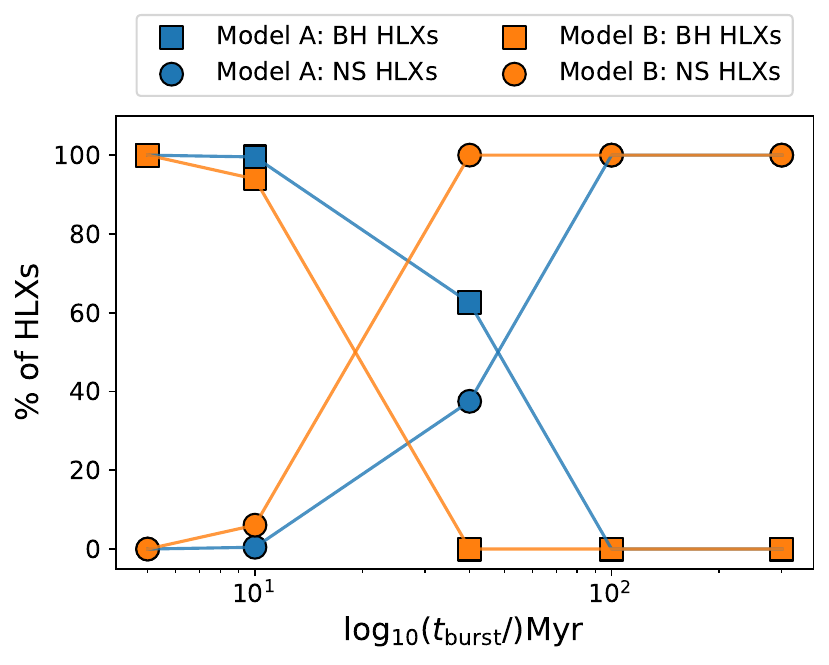}
\caption{Percentage of HLXs (having X-ray luminosities $\gtrsim 10^{41}\, \rm erg\, s^{-1}$) with maximally beamed emission (with $b = 3.2\times 10^{-3}$) for all the burst ages, distinguishing between the type of accretor (NS, shown by circles, or BH, shown by squares). The blue symbols show the results for model\,A, and the orange symbols are for model\,B.}
\label{fig:hyp_beam_v_age}
\end{figure}

\subsection{Hyperluminous X-ray sources}  

In our synthetic populations, HLXs are identified by their extremely high X-ray luminosities of $\gtrsim 10^{41}\, \rm erg\, s^{-1}$, which is much brighter than typical ULXs. Even though most observed HLXs are believed to be accreting intermediate-mass BHs \citep{2011AN....332..392F, 2011ApJ...743....6S, 2011ApJ...735...89L, 2011ApJ...734..111D, 2015ApJ...811...23Y}, it is worth exploring extreme luminosities in the context of accreting NSs and stellar-mass BHs. \citet{2015ApJ...810...20W} found a large contribution from NS ULXs to HLXs, as accreting NSs would tend to exhibit stronger geometrical beaming (their Eddington luminosity is much lower compared to BHs). In our ULX populations, all the ULXs reaching these extreme luminosities are geometrically beamed (having $b<1$). However, the amount of beaming differs at different burst ages. 

We refer to Figures\,\ref{fig:combined_xlf_m64} and \ref{fig:combined_xlf_m44} to study the kinds of binaries that appear as HLXs in our \texttt{POSYDON} populations. In these figures, we also show the fraction of the different accretor types that received the strongest beaming (with $b=0.0032$; our lower-threshold for $b$) for our populations of HLXs in Figure\,\ref{fig:hyp_beam_v_age}. For young populations at burst ages 5\,Myr and 10\,Myr, BH XRBs with massive H-rich donors undergoing RLO easily reach extreme luminosities via geometrical beamed emission. Since the population at 5\,Myr only has BH accretors, all the strongest beamed HLXs are BH accreting binaries. At 10\,Myr, even though the HLXs are still dominated by BH XRBs, HLXs with the strongest geometrical beaming begin to include some NS ULX systems ($\lesssim 5\%$). At 40\,Myr, different populations increase to comprise the HLX population, including wind-fed BH XRBs and RLO NS XRBs, all of which have H-rich donors (model\,B does not have the population of wind-fed BH ULXs present in model\,A). At this age in model\,B, the strongly beamed HLXs are fully dominated by the NS ULXs (at $100\%$ of the HLXs), as strong kicks have reduced the number of surviving BH binaries. All the strongly beamed HLXs in the older populations are fully dominated by NS accretors with different types of donors (H-rich or He-rich).

\section{Discussion and conclusions}
\label{sec:disc_and_conclusions}
To summarize, we investigated the effect of stellar age on populations of ULXs by generating various populations at fixed burst ages. We used \texttt{POSYDON}, a new binary population synthesis code that utilizes detailed calculations of binary interactions, to generate populations of ULXs at different burst ages, following two of the best-fitting models from \citet{2023A&A...672A..99M}. The main differences between the two models are listed in Table\,\ref{table:c4_best_fit}. We ran simulations of $10^7$ ZAMS binaries per population, each population corresponding to burst ages of 5\,Myr, 10\,Myr, 40\,Myr, 100\,Myr, 300\,Myr, and 1000\,Myr. Using the comparison of the resulting ULX populations, we reached the following key conclusions:
\begin{itemize}
    \item The treatment of the BH kicks greatly affects the total amount of XRBs observed. The higher kicks in model\,B (where BH kicks are not normalized) led to a higher disruption of binaries, reducing the overall binaries with ULX-like luminosities (decrease of up to seven times at the starburst age of 40\,Myr). As this affects the demographic of the populations, it also results in a relative dominance of NS ULXs in populations with stronger kicks (model\,B).
    \item We find that geometrically beamed emission is affected by the burst age as well as the BH kick prescription, and on average, less than $50\%$ of BH ULXs and all of NS ULXs are beamed. Populations with stronger kicks (where NSs dominate) have a higher fraction of ULXs with beamed emission.
    \item There is an inverse correlation between the number of ULXs and the age of the burst, with the largest number of ULXs at 5\,Myr and none at 1000\,Myr. We find that younger ULX populations ($\lesssim 10\,\rm Myr$) are dominated by HMXBs, populations with ages $\sim 40\,\rm Myr$ are dominated by IMXBs, and older populations (100 to 300\,Myr) are dominated by LMXBs. 
    \item As the populations age, the ratio of BH to NS accretors decreases, which affects the other orbital properties of ULXs. The type of ULX donor changes, with younger populations ($< 100\,\rm Myr$) having massive H-rich MS donors, while older populations ($> 100\,\rm Myr$) are dominated by low-mass H-rich post-MS donors. At 100\,Myr, stripped He-rich donors are prominent in ULXs.
    \item Accretion plays a significant role in the observation of pulsating ULXs, affecting the fraction of NS ULXs inferred in a population. Depending on the radius at which the accretion disk begins (at the NS radius or the magnetospheric radius), the number of NS ULXs where pulses are observed varies (from $30\%$ to none of the total NS ULX population depending on the stellar age), implying many more NS ULXs might be present in the ULX populations compared to those that show coherent pulsations. Additionally, NS ULXs with strong magnetic fields ($\gtrsim 10^{12}\rm\,G$) are increasingly difficult to observe.
    \item Hyperluminous X-ray sources (ULXs with luminosities $\gtrsim 10^{41}\, \rm erg\, s^{-1}$) are present at all the burst ages explored (except at 1000\,Myr), with an inverse correlation with age. All the HLXs, at all ages, have geometrically beamed emissions.
\end{itemize}

In addition to the effects of age on ULX populations, we also observed a reflection of the physical assumptions made about binary evolution. We find that BHs dominate ULXs in younger starburst populations, and NSs start to occupy more space in the demographic after about 100\,Myr, confirming the results of \citet{2017ApJ...846...17W}. \citet{2017ApJ...846...17W} also found that the production of BH ULXs is limited to up to 200\,Myr. Similarly, in our \texttt{POSYDON} burst populations, BH ULXs appear before 100\,Myr (dominating the populations up to 40\,Myr), after which NS ULXs are the only main contributors, with negligible contributions from BH ULXs. The specific contribution from BH ULXs also depends on the natal kicks because model\,A (with weaker BH kicks) has more BH accretors than model\,B (see Figure\,\ref{fig:bh_total_v_age} and Section\,\ref{sec:results:acc_nat}). Since we generated ULX populations from star-forming bursts of specific ages, the results presented in this study are not directly comparable to observations, but the populations can be convolved with star-formation histories to do so.

Our results regarding the geometrically beamed emission in ULXs are similar to those presented by \citet{2019ApJ...875...53W}. They found that while most NS ULXs are beamed whenever they are present, the fraction of BH ULXs that have beamed emission decreases with age. This also corroborates the work by \citet{2016MNRAS.458L..10K}, where they suggested that NS ULXs are more likely to be beamed compared to BH ULXs due to the lower NS Eddington limit. However, the calculations involved are heavily dependent on the simplistic model used to  define X-ray luminosity from super-Eddington accretion disks (described in Section\,\ref{sec:methods:x-ray_lum}) and might show different results with a different model. \citet{2020MNRAS.494.3611K} introduced the idea that many pulsating ULXs have a strong misalignment between the NS spin axis and the accretion disk axis, which enables the X-ray pulses to escape and be observed. As the NS accretes matter and hence angular momentum, the axes align and X-ray pulses are further suppressed. Generally, we found in our study that X-ray pulses are suppressed for 70\%\ to 100\% of the NS ULXs, depending on the population age and assuming a certain accretion model.

% metallicity
While the ULX populations were generated at solar metallicity, ULXs have been observed in excess in low-metallicity regions compared to solar metallicity, when normalized by the SFR \citep{2005MNRAS.356...12S, 2007Ap&SS.311..213S,2009MNRAS.395L..71M,2010MNRAS.408..234M,2010ApJ...725.1984L, 2011ApJ...741...10K,2013ApJ...769...92P,2016ApJ...818..140B, 2020MNRAS.498.4790K}. This behavior is attributed to the fact that at lower metallicities, stars have more compact and massive stellar cores \citep{2009MNRAS.395L..71M,2009MNRAS.400..677Z} that can initiate RLO in close orbits, leading to a higher production of HMXBs per SFR \citep{2004ApJ...609..133M,2006MNRAS.370.2079D,2007Ap&SS.311..213S,2009MNRAS.395L..71M,2010ApJ...725.1984L, 2013ApJ...764...41F, 2013ApJ...776L..31F, 2016ApJ...818..140B}, which has been reproduced by theoretical studies \citep{2014MNRAS.437.1187Z, 2017ApJ...846...17W}. In the future, we will carry out a study of ULX populations with detailed population synthesis techniques, as used in this work, at low-metallicities, as it will further our understanding of ULXs in galaxies where the aforementioned excess is observed, as well as in the high-redshift Universe.

This study was carried out to observe the effects of age on ULX populations and the general trends in the populations while also investigating the effects of certain model assumptions. We find that while the populations overall followed similar trends between different models of physical assumptions, comparative studies between observations and simulations could shed light on certain physical processes. Additionally, extending this study to include a wide range of metallicities and star-forming histories would be the next step in answering many questions related to ULXs and the field of binary evolution as a whole. 

\begin{acknowledgements}
%The authors thank the anonymous referee for their constructive comments that helped improve the manuscript. 

The authors thank the anonymous referee for their con- structive comments that helped improve the manuscript. The POSYDON project is supported primarily by two sources: the Swiss National Science Foundation (PI Fragos, project numbers PP00P2\_211006 and CRSII5\_213497) and the Gordon and Betty Moore Foundation (PI Kalogera, grant award GBMF8477). DM acknowledges support from the Swiss National Science Foundation (project number PP00P2\_176868) and the European Research Council (ERC) under the European Union’s Horizon 2020 research and innovation programme (grant agreement No. 101002352). KK acknowledges support from the Federal Commission for Scholarships for Foreign Students for the Swiss Government Excellence Scholarship (ESKAS No. 2021.0277), and the Spanish State Research Agency, through the María de Maeztu Program for Centers and Units of Excellence in R\&D, No. CEX2020-001058-M. S.S.B., T.F., and Z.X. were supported by the project number PP00P2\_211006. S.S.B. was also supported by the project number CRSII5\_213497. Z.X. acknowledges support from the Chinese Scholarship Council (CSC). A.D., K.A.R., P.M.S. and M.S. were supported by the project number GBMF8477. EZ acknowledges funding support from the European Research Council (ERC) under the European Union’s Horizon 2020 research and innovation programme (Grant agreement No. 772086) as well as from the Hellenic Foundation for Research and Innovation (H.F.R.I.) under the “3rd Call for H.F.R.I. Research Projects to support Post-Doctoral Researchers” (Project No: 7933). The computations were performed at Northwestern University on the Trident computer cluster (funded by the GBMF8477 award) and at the University of Geneva on the Yggdrasil computer cluster.

\end{acknowledgements}

\bibliographystyle{aa}
\bibliography{main.bib}

\newpage\hbox{}\newpage
\begin{appendix}
%\label{sec:appendix}
\section{Synthetic XLFs for model~B}\label{sec:appendix}

\begin{figure*}[htb]
\hspace{2.0cm}\includegraphics[width=0.7\textwidth]{legend_type.png}\\
\centering
\begin{tikzpicture}
\node (img1)  {\includegraphics[width=0.95\textwidth]{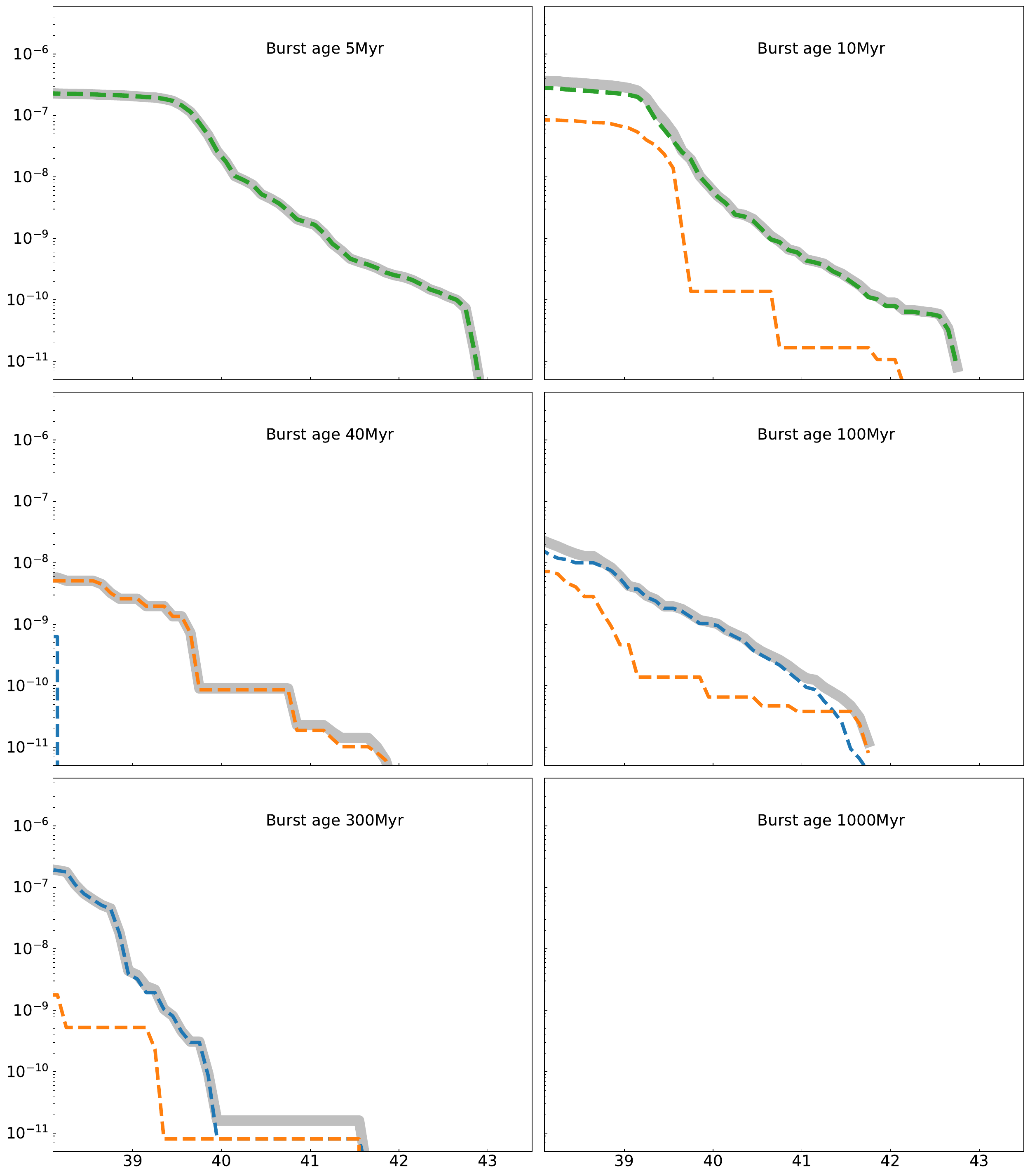}};
\node[node distance=0cm, yshift=-10.5cm] {log$_{10} (L_{\rm X}/\mathrm{erg\,s^{-1}}$)};
\node[node distance=0cm, rotate=90, anchor=center,yshift=9.0cm]  {N$(>L_{\rm X})$ / $\mathrm{M_\odot}$};
\end{tikzpicture}
\caption{Synthetic XLFs of burst populations for model\,B. The panels show the different types of XRBs, namely, HMXBs, IMXBs, and LMXBs, as described in the legend. The combined XLF is shown by the dotted-gray line.}
\label{fig:combined_type_m44}
\end{figure*}

\begin{figure*}[htb]
\hspace{2.0cm}\includegraphics[width=0.7\textwidth]{legend.png}\\
\centering
\begin{tikzpicture}
\node (img1)  {\includegraphics[width=0.95\textwidth]{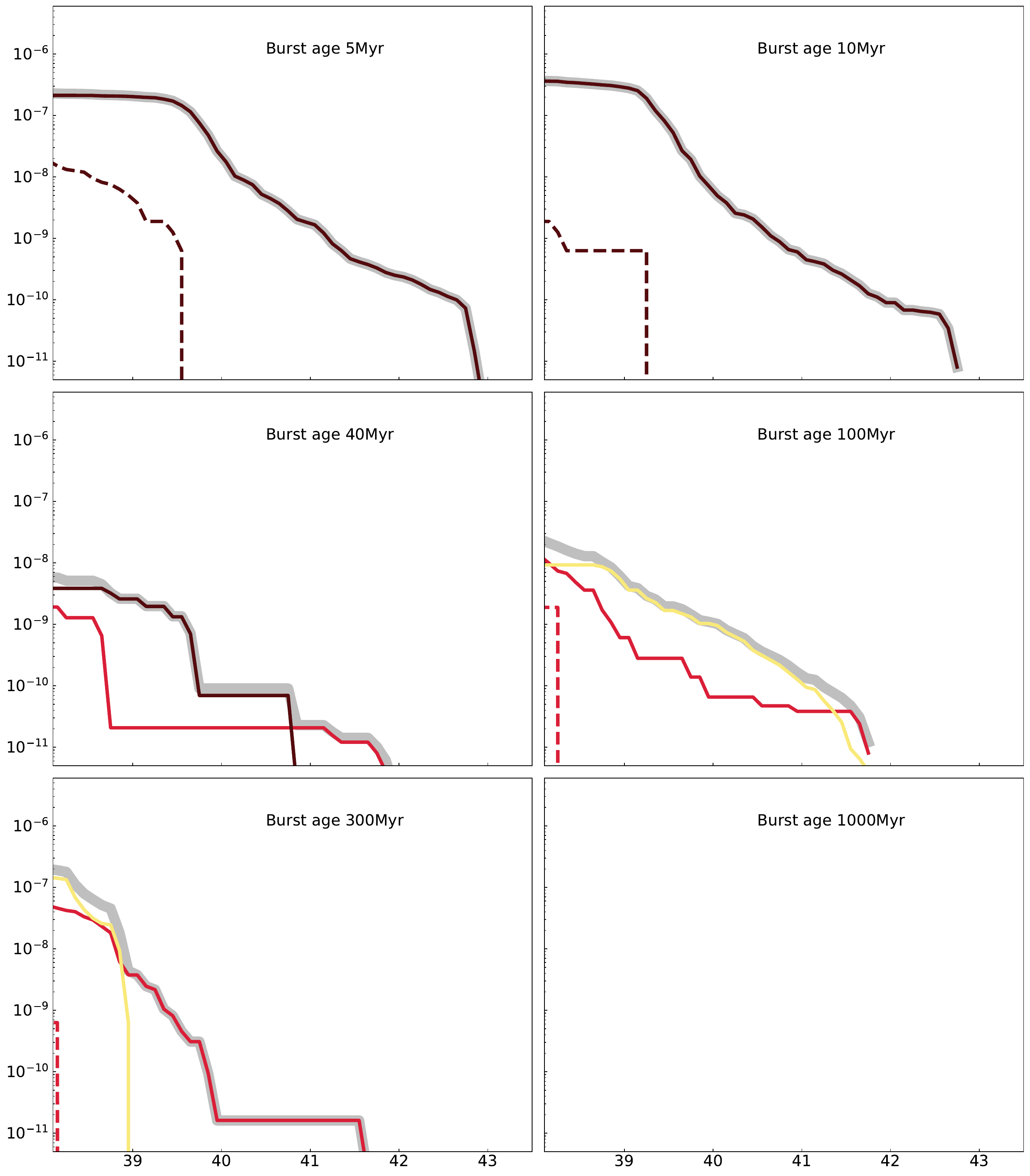}};
\node[node distance=0cm, yshift=-10.5cm] {log$_{10} (L_{\rm X}/\mathrm{erg\,s^{-1}}$)};
\node[node distance=0cm, rotate=90, anchor=center,yshift=9.0cm]  {N$(>L_{\rm X})$ / $\mathrm{M_\odot}$};
\end{tikzpicture}
\caption{Synthetic XLFs of burst populations for model\,B. The populations are further split into the type of the XRBs (RLO or wind), type of accretors (BH or NS), and type of donors (H-rich or He-rich). The combined XLF is shown by the dotted-gray line.}
\label{fig:combined_xlf_m44}
\end{figure*}

We present the figures describing the populations for model\,B. In Figure\,\ref{fig:combined_type_m44}, we show the synthetic XLFs of burst populations at different burst ages divided into sub-populations of different types of XRBs (namely, HMXBs, IMXBs, and LMXBs; similar to Figure\,\ref{fig:combined_type_m64}). Figure\,\ref{fig:combined_xlf_m44} further splits the populations by type of accretion (RLO or wind), accretor (BH or NS), and donor (H-rich or He-rich).

%%%%%%%%%

%%%%%%%%%%%%%%%%%%%

\end{appendix}

\end{document}